\documentclass{elsarticle}
				
\usepackage[a4paper]{geometry} % change page margins
\usepackage[utf8]{inputenc}
\usepackage{amsmath}
\usepackage{amsthm}
\usepackage{empheq}
\usepackage{bm} % bold math

\usepackage{stix2}

\usepackage{graphicx}
\usepackage{epstopdf} 
\usepackage{multirow}
\usepackage{tocbasic}
\usepackage{caption}
\usepackage{subcaption}
\usepackage{placeins}%for float barrier
\usepackage{siunitx}
\usepackage{booktabs} % professional tables
\usepackage{todonotes} % todo's
\usepackage{nameref}
\usepackage[mathlines]{lineno}
\usepackage{etoolbox}

\graphicspath{{figures/}}
\usepackage{hyperref} % references become links

\newcommand{\vt}[1]{\bm{#1}}
\newcommand{\rd}{\textrm{d}}

\newcommand{\dd}[2]{\frac{\partial #1}{\partial #2}}
\newcommand{\ul}[1]{\underline{#1}}

\newcommand{\hl}[1]{\textcolor{black}{#1}}

\newcommand{\R}[1]{\label{#1}\label{#1}}

\begin{document}
% start line numbers
%\linenumbers

\title{A novel pressure-free two-fluid model for one-dimensional incompressible multiphase flow}              % mandatory field
\author[add1]{B. Sanderse}
\ead{B.Sanderse@cwi.nl, corresponding author}
\author[add1,add3]{J.F.H. Buist}
\author[add2,add3]{R.A.W.M. Henkes}

\address[add1]{Centrum Wiskunde \& Informatica (CWI), Amsterdam, The Netherlands}
\address[add2]{Shell Technology Centre Amsterdam, Amsterdam, The Netherlands}
\address[add3]{Delft University of Technology, Delft, The Netherlands}

\begin{abstract}
A novel pressure-free two-fluid model formulation is proposed for the simulation of one-dimensional incompressible multiphase flow in pipelines and channels. The model is obtained by simultaneously eliminating the volume constraint and the pressure from the widely used two-fluid model (TFM). The resulting `pressure-free two-fluid model' (PF-TFM) has a number of attractive features: (i) it features four evolution equations (without additional constraints) that can be solved very quickly with explicit time integration methods; (ii) it keeps the conservation properties of the original two-fluid model, and therefore the correct shock relations in case of discontinuities; (iii) its solutions satisfy the two TFM constraints exactly: the volume constraint and the volumetric flow constraint; (iv) it offers a convenient form to analytically analyse the equation system, since the constraint has been removed. \\
A staggered-grid spatial discretization and an explicit Runge-Kutta time integration method are proposed, which keep the constraints exactly satisfied when numerically solving the PF-TFM. Furthermore, for the case of strongly imposed boundary conditions, a novel adapted Runge-Kutta formulation is proposed that keeps the volumetric flow exact in time while retaining high order accuracy. Several test cases confirm the theoretical properties and show the efficiency of the new pressure-free model.
\end{abstract}


%\date{\today}

\maketitle

%\tableofcontents

%%%%%%%%%%%%%%%%%%%%%%%%%%%%%%%%%%%%%%%%%%%%%%%%%%%%%%%%%
%%%%%%%%%%%%%%%%%      Main matter     %%%%%%%%%%%%%%%%%%
%%%%%%%%%%%%%%%%%%%%%%%%%%%%%%%%%%%%%%%%%%%%%%%%%%%%%%%%%

\section{Introduction}\label{sec:introduction}
%\subsection{Background}
The two-fluid model (TFM) for one-dimensional multiphase flow is an important model to study, for example, the behaviour of oil and gas transport in long pipelines. In this article we study its incompressible simplification. \R{com3_1}\hl{An example application is flushing (cleaning) an oil-filled pipeline at atmospheric conditions with water, in which case the incompressible assumption is accurate.} Like in single-phase flow models, the incompressibility assumption necessitates the use of careful space- and time-discretization methods. In the incompressible TFM, the main difficulty in defining accurate numerics arises from the presence of the volume constraint (the two phases should together fill the pipeline), its associated volumetric flow constraint (the mixture velocity field should be divergence-free), and the role of the pressure. 

One approach for numerically solving the incompressible TFM is to eliminate the pressure from the four-equation system (+ 1 constraint) and to rewrite this system into a two-equation system (+ 2 constraints). This leads to the `no-pressure wave' model or the `fixed-flux' model (FFM) suggested by Watson \cite{Watson1989}, and used for example in \cite{Akselsen2016,Holmas2008,LopezdeBertodano2017,Omgba-Essama2004}. Another two-equation model is the `reversed density' model developed by Keyfitz et al.\ \cite{Keyfitz2003} and employed for example in \cite{Wangensteen2010}. In these models, the pressure is generally computed as a post-processing step. A general problem with these two-equation systems is that they are only valid for smooth solutions. This means that in the presence of shocks different jump conditions are obtained than for the four-equation model \cite{Akselsen2016} (an example of shock waves in the two-fluid model are roll waves \cite{Watson1989}). Furthermore, the fixed-flux assumption often limits these studies to stationary boundary conditions. 

Another approach is to stick to the four-equation formulation and keep the pressure, and to use a pressure-correction method. A pressure equation is then typically obtained by substituting the momentum equations in the combined mass conservation equation, either with employing the volume constraint equation \cite{Liao2008}, or without employing the volume constraint equation \cite{Issa2003,LopezdeBertodano2017,Wang2015}. In our recent work \cite{Sanderse2018} we proposed a constraint-consistent pressure equation, which was shown to be both a generalized Riemann invariant of the continuous equations, and a hidden constraint of the semi-discrete differential algebraic equation system. The proposed Runge-Kutta time integration method was shown to be high-order accurate while satisfying both the volume constraint and the volumetric flow constraint. However, this approach is computationally relatively costly, as one needs to solve a pressure Poisson equation at each stage of the Runge-Kutta method, with the additional difficulty that the Poisson matrix depends on the actual solution and therefore changes in time. %, so one cannot precompute its LU-decomposition.

In this paper we propose a novel incompressible two-fluid model formulation in which the pressure and the volume constraint have been eliminated. The elimination process will be such that the resulting pressure-free model still satisfies both the volume constraint and the volumetric flow constraint. Compared to the four-equation approach, our novel formulation is much faster to evaluate since it does not require the solution of a Poisson equation. Compared to the two-equation approach, the new model keeps the correct shock solutions, and furthermore it does not rely on the assumption of the `fixed-flux'. 

%Since we use the four-equation `conservative' form\footnote{We note that the momentum equations in the four-equation model are not fully conservative due to the presence of the hold-up fractions in the pressure gradient. This is a topic of study in itself, and we refer the interested reader to \cite{Toumi1996}.}, 
The outline of this paper is as follows: in section \ref{sec:governing_equations} the incompressible two-fluid model equations are given, in section \ref{sec:PFTFM} the new pressure-free model is derived, and in sections \ref{sec:spatial_discretization} and \ref{sec:time_discr} appropriate spatial and temporal discretizations are proposed. Results are shown in section \ref{sec:results}.

\section{Governing equations for incompressible two-phase flow in pipelines}\label{sec:governing_equations}
The incompressible two-fluid model can be derived by considering the stratified flow of liquid and gas in a pipeline (for a recent discussion of the two-fluid model, see for example \cite{LopezdeBertodano2017}). The main assumptions that we make are that the flow is one-dimensional, stratified, incompressible, and isothermal. The transverse pressure variation is introduced via level gradient terms. Surface tension is neglected. This leads to the following system of equations, which we call the `original' TFM \cite{Sanderse2018}:
\begin{equation}\label{eqn:TFM}
\frac{\partial \vt{U}}{\partial t} + \frac{\partial \vt{f}(\vt{U})}{\partial s} + \vt{h}(\vt{U}) \dd{p}{s} + \vt{S}(\vt{U}) = 0,
\end{equation}
where
\begin{equation}
\vt{U} = \begin{pmatrix}
U_{1} \\ U_{2} \\ U_{3} \\ U_{4}
\end{pmatrix} = \begin{pmatrix}
\rho_{g} A_{g} \\ \rho_{l} A_{l} \\ \rho_{g} u_{g} A_{g} \\ \rho_{l} u_{l} A_{l} 
\end{pmatrix}, \qquad 
\vt{f} (\vt{U}) = 
\begin{pmatrix}
\rho_{g} u_{g} A_{g} \\  \rho_{l} u_{l} A_{l} \\ \rho_{g} u_{g}^2 A_{g} + K_{g} \\ \rho_{l} u_{l}^2 A_{l}  + K_{l}
\end{pmatrix}, \qquad
\vt{h}(\vt{U}) =
\begin{pmatrix}
0 \\
0 \\
A_{g} \\
A_{l}
\end{pmatrix}, \qquad
\vt{S}(\vt{U}) =
\begin{pmatrix}
0 \\
0 \\
S_{g} \\
S_{l}
\end{pmatrix},
\end{equation}
%
%\begin{align}
%\frac{\partial}{\partial t} \left( \rho_{g} A_{g} \right) + \frac{\partial }{\partial s} \left( \rho_{g} u_{g} A_{g} \right) &= 0, \label{eqn:conservation_mass_gas} \\
%\frac{\partial}{\partial t} \left( \rho_{l} A_{l} \right) + \frac{\partial }{\partial s} \left( \rho_{l} u_{l} A_{l} \right) &= 0,\label{eqn:conservation_mass_liq}\\
%\frac{\partial }{\partial t} \left( \rho_{g} u_{g} A_{g} \right) + \frac{\partial }{\partial s} (\rho_{g} u_{g}^2 A_{g}) &= -\frac{\partial p }{\partial s} A_{g} + LG_{g} + \underbrace{ - \tau_{gl} P_{gl} - \tau_{g} P_{g} - \rho_{g} A_{g} g_s +  F_{\text{body}} {A_g}}_{S_{g}},\\
%\frac{\partial }{\partial t} \left( \rho_{l} u_{l} A_{l} \right) + \frac{\partial }{\partial s} (\rho_{l} u_{l}^2 A_{l}) &= -\frac{\partial p }{\partial s} A_{l} + LG_{l} + \underbrace{\tau_{gl} P_{gl} - \tau_{l} P_{l} - \rho_{l} A_{l} g_s +  F_{\text{body}} {A_l}}_{S_{l}},\label{eqn:conservation_momentum}
%\end{align}
supplemented with the \textit{volume constraint equation}:
\begin{equation}\label{eqn:volume_constraint}
A_g + A_l - A =0.
\end{equation}
In these equations the subscript denotes either gas ($g$) or liquid ($l$). The model features four evolution equations, one constraint equation, and five unknowns ($\vt{U}$ and $p$), which are a function of the independent variables $s$ (coordinate along the pipeline axis) and $t$ (time). The primitive variables $A_{g}$, $u_{g}$ etc.\ can be expressed fully in terms of $\vt{U}$, e.g.\
\begin{equation}\label{eqn:Ag}
A_{g} = \frac{U_{1}}{\rho_{g}}, \qquad u_{g} = \frac{U_{3}}{U_{1}}.
\end{equation}
$\rho$ denotes the density (assumed constant), $A$ the cross-sectional area of the pipe, $A_{g}$ and $A_{l}$ (also referred to as the hold-ups) the cross-sectional areas occupied by the gas or liquid, $R$ the pipe radius, $h$ the height of the liquid layer measured from the bottom of the pipe, $u$ the phase velocity, $p$ the pressure at the interface, $\tau$ the shear stress (with the wall or at the interface), $g$ the gravitational constant, $\varphi$ the local upward inclination of the pipeline with respect to the horizontal, and the components of gravity are $g_n=g \cos \varphi$ and $g_s = g \sin \varphi$. See figure \ref{fig:stratified_flow}.

\begingroup
\begin{figure}[hbtp]
\fontfamily{lmss}
\fontsize{10pt}{12pt}\selectfont
\centering 
\def\svgwidth{0.8 \textwidth}
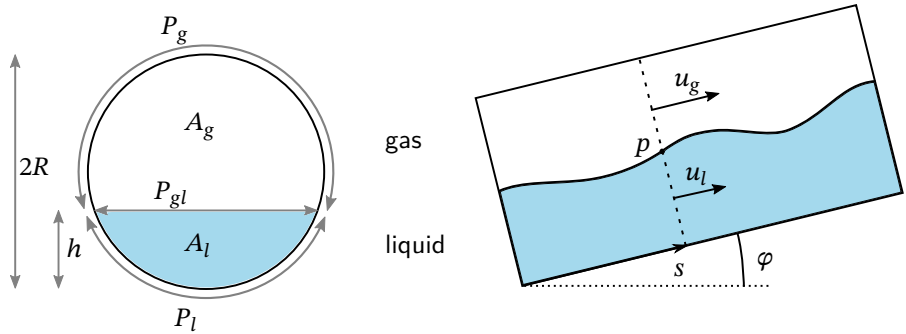 
\caption{Stratified flow in a pipeline. Left: cross-sectional view; right: side view.}
\label{fig:stratified_flow}
\end{figure}
\endgroup

The level gradient terms $K$ for a pipe (compressible or incompressible flow) are given by \cite{Sanderse2017}:
\begin{align}
K_{g} = - \rho_{g} g_n \left[ (R-h) A_{g} + \frac{1}{12} P_{gl}^3\right], \label{eqn:compressible_HG}\\
K_{l} = - \rho_{l} g_n \left[ (R-h) A_{l} - \frac{1}{12} P_{gl}^3\right],
\end{align}
while for a channel (compressible or incompressible) one has:
\begin{align}
K_{g} &= - \frac{1}{2} \rho_{g} g_{n}  A_{g}^2, \\
K_{l } &= \frac{1}{2} \rho_{l} g_{n} A_{l}^2.
\end{align} 
In a pipe geometry, $A_l$ and $h$ are related by a non-linear algebraic expression, see e.g.\ \cite{Sanderse2018}; in a channel one has $A_{l}=h$ (assuming unit depth). The wetted and interfacial perimeters $P_{g}$, $P_{l}$ and $P_{gl}$ can be expressed in terms of the liquid hold-up $A_{l}$ or the interface height $h$ (see \cite{Sanderse2018}). 
%We will make use of Biberg's approximation \cite{Biberg1999} to avoid the solution of a non-linear equation. 

The source terms $S_{g}$ and $S_{l}$ constitute friction, gravity, and driving pressure gradient, given by
\begin{align}
S_{g} &= \tau_{gl} P_{gl} + \tau_{g} P_{g} + \rho_{g} A_{g} g_s + A_{g} \frac{\rd p_{0}}{\rd s}, \\
S_{l} &= -\tau_{gl} P_{gl} + \tau_{l} P_{l} + \rho_{l} A_{l} g_s + A_{l} \frac{\rd p_{0}}{\rd s}.
\end{align}
%The body force $F_{\text{body}}$ in the gas and liquid momentum equations is for example a driving pressure force for the simulations that involve periodic boundary conditions, or a source term to force an analytical solution (see \ref{sec:MMS}). 
The friction models are described in \cite{Sanderse2018}. The source terms $S_{g}$ and $S_{l}$ do not contain spatial or temporal derivatives of the unknowns of the system ($\vt{U}$ and $p$). The driving pressure gradient $\frac{\rd p_{0}}{\rd s}$ is imposed in case of periodic boundary conditions.

Initial and boundary conditions determine whether the equations form a well-posed initial boundary value problem. It is well-known that this is not always the case: the two-fluid model equations can become ill-posed depending on the velocity difference between the phases \cite{Barnea1994,Lyczkowski1978,Stewart1984}. In this paper we will restrict ourselves to test cases for which the model is well-posed. Furthermore, as will become clear in the next section, we require the initial conditions to be consistent, meaning that at $t=0$ the solution should not only satisfy the volume constraint but also the \textit{volumetric flow constraint}:
\begin{align}
%A_{g} + A_{l} - A &= 0, \label{eqn:IC1} \\
\dd{}{s} \left( u_{g} A_{g} + u_{l} A_{l}  \right) &= 0. \label{eqn:IC2}
\end{align}

\section{A new pressure-free two-fluid model}\label{sec:PFTFM}
\subsection{The hidden constraints of the incompressible TFM}
In this section we show how the pressure and the constraint can be removed from the incompressible TFM. For this purpose, we use the characteristic analysis of the incompressible two-fluid model proposed in \cite{Sanderse2018}. In that work, it was shown that the volume constraint, equation \eqref{eqn:volume_constraint}, could be rewritten in two alternative forms. 

The first form is the volumetric flow constraint, obtained by premultiplying the TFM equations with 
\begin{equation}
\vt{l}_{3} = \begin{pmatrix}
\frac{1}{\rho_{g}} & \frac{1}{\rho_{l}} & 0 & 0
\end{pmatrix},
\end{equation}
leading to
\begin{equation}\label{eqn:l3}
\begin{split}%
\qquad \qquad \vt{l}_{3} \cdot \left( \frac{\partial \vt{U}}{\partial t} + \frac{\partial \vt{f}(\vt{U})}{\partial s} + \vt{h}(\vt{U}) \dd{p}{s} + \vt{S}(\vt{U})  \right) = 0, \\
\Rightarrow \dd{A_{g}}{t} + \dd{A_{l}}{t}+ \frac{\partial }{\partial s} \left( A_{g} u_{g} + A_{l} u_{l} \right) = 0, \\
 \Rightarrow \frac{\partial }{\partial s} \left( A_{g} u_{g} + A_{l} u_{l} \right) = 0. 
\end{split}
\end{equation}
The last equation is the \textit{volumetric flow constraint}. Note that in this derivation we used the time-derivative of the volume constraint,
\begin{equation}\label{eqn:constraint_ddt}
\dd{}{t} \left( A_{g} + A_{l} \right) = 0,
\end{equation}
so the volumetric flow constraint can be interpreted as a combination of mass equations and volume constraint. Integrating the volumetric flow constraint in space gives
\begin{equation}\label{eqn:constraint_diff1}
A_{g} u_{g} + A_{l} u_{l} = Q(t).
\end{equation}

The second form is the pressure equation, obtained by premultiplying the TFM equations with 
\begin{equation}
\vt{l}_{4} = \begin{pmatrix}
0 & 0 & \frac{1}{\rho_{g}} & \frac{1}{\rho_{l}} 
\end{pmatrix},
\end{equation}
leading to 
\begin{multline}\label{eqn:l4}
\vt{l}_{4} \cdot \left( \frac{\partial \vt{U}}{\partial t} + \frac{\partial \vt{f}(\vt{U})}{\partial s} + \vt{h}(\vt{U}) \dd{p}{s} + \vt{S}(\vt{U})  \right) = 0, \\   
\Rightarrow \dd{}{t} \left(u_{g} A_{g} + u_{l} A_{l} \right) + \dd{}{s} \left(u_{g}^2 A_{g} + \frac{K_{g}}{\rho_{g}} + u_{l}^2 A_{l} + \frac{K_{l}}{\rho_{l}} \right) +  \left(\frac{A_{l}}{\rho_{l}} + \frac{A_{g}}{\rho_{g}}\right) \dd{p}{s}  +  \frac{S_{g}}{\rho_{g}}  + \frac{S_{l}}{\rho_{l}} = 0,\\
\Rightarrow - \left(\frac{A_{l}}{\rho_{l}} + \frac{A_{g}}{\rho_{g}}\right) \dd{p}{s}  = \dot{Q}(t) + \dd{}{s} \left(u_{g}^2 A_{g} + \frac{K_{g}}{\rho_{g}} + u_{l}^2 A_{l} + \frac{K_{l}}{\rho_{l}} \right) +   \frac{S_{g}}{\rho_{g}}  + \frac{S_{l}}{\rho_{l}}. 
\end{multline}
At the heart of this derivation is the time-derivative of equation \eqref{eqn:constraint_diff1}, which in itself was obtained from the time-derivative of the volume constraint, see equation \eqref{eqn:constraint_ddt}. In other words, equation  \eqref{eqn:l4} can be interpreted as an equation for the second-order time-derivative of the constraint:
\begin{equation}\label{eqn:constraint_diff2}
\dd{^2}{t^2} \left( A_{g} + A_{l} \right) = 0.
\end{equation}
Note that the pressure gradient equation \eqref{eqn:l4} is different from the expression commonly used in literature (see e.g.\ \cite{Holmas2010}), in which the gas and liquid momentum equations are added \textit{without scaling by their respective densities}. Although such a pressure equation can be useful for postprocessing computations, it cannot be used to replace the volume constraint, because the volume constraint has not been employed in its derivation. In contrast, our derived pressure equation \eqref{eqn:l4} is a combination of the momentum equations, mass equations, \textit{and} the volume constraint.

\subsection{The pressure-free TFM}\label{sec:PFTFM_derivation}
The key insight of this paper is the following. The expression for the pressure gradient, equation \eqref{eqn:l4}, can be substituted back into the TFM equations. For this purpose, first write the pressure equation in short notation as
%We split the source term $\vt{S}(\vt{U},p)$ in a term involving the pressure and the remaining terms ($S_{g}$ and $S_{l}$):
%\begin{equation}
%\vt{S}(\vt{U},p) = \vt{S}_{p} (\vt{U},p) + \vt{S}_{u} (\vt{U}),
%\end{equation}
%so that we have
\begin{equation}\label{eqn:dpds}
\dd{p}{s} =  - \frac{\rho_{l} \rho_{g}}{\hat{\rho}} \vt{l}_{4} \cdot \left( \frac{\partial \vt{f}(\vt{U})}{\partial s} + \vt{S}(\vt{U}) \right) -  \frac{\rho_{l} \rho_{g}}{\hat{\rho}} \dot{Q}(t),
\end{equation}
where
\begin{equation}
\hat{\rho}(\vt{U}) := \rho_{g} A_{l} + \rho_{l} A_{g}.
\end{equation}
The pressure gradient term as present in the TFM then reads
\begin{equation}\label{eqn:hdpds}
\begin{split}
\vt{h} (\vt{U}) \dd{p}{s} = 
\begin{pmatrix}
0 \\ 
0 \\
A_{g} \\
A_{l} 
\end{pmatrix} 
\dd{p}{s}
&=-
\begin{pmatrix}
0 \\ 
0 \\
\frac{A_{g} \rho_{l} \rho_{g}}{\hat{\rho}} \\
\frac{A_{l} \rho_{l} \rho_{g}}{\hat{\rho}}
\end{pmatrix} 
\cdot 
\vt{l}_{4} \cdot \left( \frac{\partial \vt{f}(\vt{U})}{\partial s} + \vt{S}(\vt{U}) \right) -  
\begin{pmatrix}
0 \\ 
0 \\
\frac{A_{g} \rho_{l} \rho_{g}}{\hat{\rho}} \\
\frac{A_{l} \rho_{l} \rho_{g}}{\hat{\rho}}
\end{pmatrix} 
\dot{Q}(t) \\ 
&= - 
\vt{B} (\vt{U})
 \cdot \left( \frac{\partial \vt{f}(\vt{U})}{\partial s} + \vt{S}(\vt{U}) \right) -  
 \begin{pmatrix}
0 \\ 
0 \\
\frac{A_{g} \rho_{l} \rho_{g}}{\hat{\rho}} \\
\frac{A_{l} \rho_{l} \rho_{g}}{\hat{\rho}}
\end{pmatrix} 
 \dot{Q}(t), 
\end{split}
\end{equation}
with
\begin{equation}
\vt{B}(\vt{U}) =  \begin{pmatrix}
0 \\ 
0 \\
\frac{A_{g} \rho_{l} \rho_{g}}{\hat{\rho}} \\
\frac{A_{l} \rho_{l} \rho_{g}}{\hat{\rho}}
\end{pmatrix} 
\cdot \vt{l}_{4} = 
\frac{1}{\hat{\rho}}
 \begin{pmatrix}
0 & 0 & 0 & 0 \\
0 & 0& 0& 0 \\
0 & 0 & A_{g} \rho_{l} & A_{g} \rho_{g} \\
0 & 0 & A_{l} \rho_{l} & A_{l} \rho_{g} 
\end{pmatrix}.
\end{equation}
The important conclusion is that \textit{the pressure gradient term in the TFM can be written in terms of a linear combination of the flux vector and the source terms}. 
%An alternative formulation is the one of Holmas, where the momentum equations are added without dividing by the densities:
%\begin{equation}\label{eqn:dpds}
%D \dd{p}{s}  =  -\dd{}{t} (\rho_{g} u_{g} h_{g} + \rho_{l} u_{l} h_{l}) - \dd{}{s} (\rho_{l} u_{l}^2 h_{l} + \rho_{g} u_{g}^2 h_{g} + \frac{1}{2} \rho_{l} g h_{l}^2- \frac{1}{2} \rho_{g} g h_{g}^2) + \frac{S_{g}}{\rho_{g}}  + \frac{S_{l}}{\rho_{l}}.  
%\end{equation}
%In the previous section we have shown that the ``pressure-free'' TFM is equivalent to the FFM. 
Upon substituting the expression for $\vt{h}(\vt{U}) \frac{\partial p}{\partial s}$ into the TFM, we obtain the new pressure-free TFM (PF-TFM) as
\begin{equation}\label{eqn:PFTFM}
\textbf{PF-TFM:} \qquad \qquad \boxed{\frac{\partial \vt{U}}{\partial t} + \vt{A} (\vt{U}) \frac{\partial \vt{f}(\vt{U})}{\partial s} + \vt{A} (\vt{U}) \vt{S}(\vt{U}) + \vt{c}(\vt{U}) \dot{Q}(t) = 0,}
\end{equation}
where
\begin{equation}\label{eqn:expressionAC}
\vt{A}(\vt{U}) = \vt{I} - \vt{B}(\vt{U}) = 
 \begin{pmatrix}
1 & 0 & 0 & 0 \\
0 & 1& 0& 0 \\
0 & 0 & 1 -A_{g} \rho_{l} / \hat{\rho} & - A_{g} \rho_{g} / \hat{\rho}\\
0 & 0 & -A_{l} \rho_{l} / \hat{\rho}&1- A_{l} \rho_{g} / \hat{\rho}
\end{pmatrix}, \qquad
\vt{c}(\vt{U}) =
\begin{pmatrix}
0 \\ 
0 \\
-\frac{A_{g} \rho_{l} \rho_{g}}{\hat{\rho}} \\
-\frac{A_{l} \rho_{l} \rho_{g}}{\hat{\rho}}
\end{pmatrix}.
\end{equation}
Equation \eqref{eqn:PFTFM} is a closed system of four equations in four unknowns (provided that $\dot{Q}(t)$ is known): both the constraint and the pressure have been removed. The volume constraint is `built into' this system of equations (it was used in the derivation of the pressure equation), and does not need to be provided as a fifth equation.  We note that making the model pressure-free comes at a small `price': one needs $\dot{Q}(t)$ as an input to the system of equations. In other words: removing the pressure requires the prescription of $\dot{Q}(t)$. Fortunately, $\dot{Q}(t)$ equals zero in many cases, e.g.\ in case of steady inflow conditions with prescribed liquid and gas flow rates. For unsteady inflow, e.g.\ an increasing liquid and gas production, $\dot{Q}(t)$ is known from the boundary condition specification. Only in special cases, such as periodic boundary conditions, $\dot{Q}(t)$ is unknown; we will assume it to be zero in that case and investigate by comparing PF-TFM solutions to TFM solutions to what extent this is indeed true.
%In the TFM, the term $Q(t)$ disappears in \eqref{eqn:dpds} due to an additional differentiation with respect to $s$, which yields the pressure Poisson equation. 

\subsection{Properties of the PF-TFM}
The PF-TFM shares the property of the TFM that it cannot be written in full conservative form. In the TFM this manifests itself in the presence of the hold-up fractions in front of the pressure gradient term; in the PF-TFM this enters through the presence of the $\vt{A}(\vt{U})$ matrix. Note that $\vt{A}(\vt{U})$ is a singular matrix, which implies that the system cannot be simplified by pre-multiplying with the inverse of $\vt{A}$. The eigenvalues of $\vt{A} (\vt{U}) \dd{\vt{f}(\vt{U})}{\vt{U}}$ are exactly the same as those obtained from the analysis of the original TFM in primitive variables in \cite{Sanderse2018}, see \ref{sec:eigenstructure}; as a consequence, the PF-TFM has the same characteristic velocities as the TFM.

The PF-TFM does not require the solution of a Poisson equation for the pressure, which significantly accelerates the code (this will be shown in the results section). Given a solution $\vt{U}$ to the PF-TFM, the pressure can be obtained as a post-processing step by solving equation \eqref{eqn:dpds}. An existing TFM implementation (based on a pressure Poisson equation) can easily be adapted by simply scaling the computed fluxes $\dd{\vt{f}}{s}$ using the weights from the $\vt{A}$ matrix, and adding the $\dot{Q}(t)$ term. The pressure Poisson equation can then be skipped.

Compared to the fixed-flux model (FFM), there are two important advantages. First, in contrast to the FFM, the PF-TFM retains the correct shock relations (i.e., those of the TFM). Second, the FFM is typically used with `fixed-flux' assumption $Q(t)=\text{constant}$ \cite{LopezdeBertodano2017}, whereas this assumption is not made in our model. 

Although the constraint is `built into' the pressure-free TFM, upon discretization care should be taken, because the volume constraint is not built into the PF-TFM in its original form, equation \eqref{eqn:volume_constraint}, but in a differentiated sense, namely equation \eqref{eqn:constraint_diff2}. This will be discussed next.

\section{Spatial discretization on a staggered grid to preserve constraints}\label{sec:spatial_discretization}
In the derivation of the pressure-free model, equation \eqref{eqn:PFTFM}, the following important property was silently used: the term present in the time-derivative in the momentum equations, e.g.\ $U_{3}=\rho_{g} A_{g} u_{g}$, is the same term as in the flux term in the mass equations. This property needs to be satisfied at a discrete level in order to obtain a pressure-free TFM that satisfies the volume constraint; this is achieved here by the use of a staggered grid.

The staggered grid consists of $N_{p}=N$ `pressure' and $N_{u}=N+1$ `velocity' volumes (with slight changes depending on the boundary conditions). Although the pressure is not present in the new formulation, this concept is so well-known that we stick to the terminology. The midpoints of the velocity volumes lie on the faces of the pressure volumes. The pressure, hold-up and phase masses are defined in the centre of the pressure volumes, whereas the velocity and momentum are defined in the centre of the velocity volumes. For details we refer to \cite{Sanderse2017}. The unknowns are the vector of conservative variables $\ul{U}(t)\in \mathbb{R}^{2N_p + 2N_u}$, which include the finite volume sizes (we use the notation $\ul{U}$ to distinguish from the vector $\vt{U}$ that appears in the continuous equations):
\begin{equation}
\ul{U}(t) = 
\begin{pmatrix}
\ul{U}_{1} (t) \\
\ul{U}_{2} (t) \\
\ul{U}_{3} (t) \\
\ul{U}_{4} (t)
\end{pmatrix} =
\begin{pmatrix}
[(\rho_{g} A_{g} \Delta s)_{1} \ldots (\rho_{g} A_{g} \Delta s)_{N}]^T \\
[(\rho_{l} A_{l} \Delta s)_{1} \ldots (\rho_{l} A_{l} \Delta s)_{N}]^T \\
[(\rho_{g} A_{g} u_{g} \Delta s)_{1/2} \ldots (\rho_{g} A_{g} u_{g} \Delta s)_{N+1/2}]^T \\
[(\rho_{l} A_{l} u_{l} \Delta s)_{1/2} \ldots (\rho_{l} A_{l} u_{l} \Delta s)_{N+1/2}]^T 
\end{pmatrix}.
\end{equation}
Including the volume size in the vector of unknowns leads to a clear physical interpretation (e.g.\ $\ul{U}_{1}$ has units of mass) and allows generalization to the case of time-dependent finite-volume sizes. $\ul{U}_{1}, \ul{U}_{2} \in \mathbb{R}^{N_p}$ and $\ul{U}_{3}, \ul{U}_{4} \in \mathbb{R}^{N_u}$ contain the values of mass and momentum at the mass and velocity volumes, respectively, and are only a function of time. With these vectors, the expression for the gas hold-up and gas velocity is (similar to \eqref{eqn:Ag}):
\begin{equation}\label{eqn:primitive_vars}
\ul{A}_{g} (t) = \frac{\Omega_{p}^{-1} \ul{U}_{1}(t)}{\rho_{g}}, \qquad \ul{u}_{g} (t) = \frac{\Omega_{u}^{-1} \ul{U}_{3}(t)}{I_{p} 
\Omega_{p}^{-1} \ul{U}_{1}(t)},
\end{equation}
where $\Omega_{p} \in \mathbb{R}^{N_p \times N_p}$ is a diagonal matrix that contains the pressure volume sizes, $\Omega_{u} \in \mathbb{R}^{N_u \times N_u}$ a diagonal matrix containing the velocity volume sizes, and $I_p \in \mathbb{R}^{N_u \times N_p}$ an interpolation matrix from pressure to velocity points.
%Together with the pressure and the volume constraint, this gives 5 unknowns for 5 equations.

\begingroup
\begin{figure}[hbtp]
\fontsize{10pt}{12pt}\selectfont
\centering 
\def\svgwidth{ \textwidth}
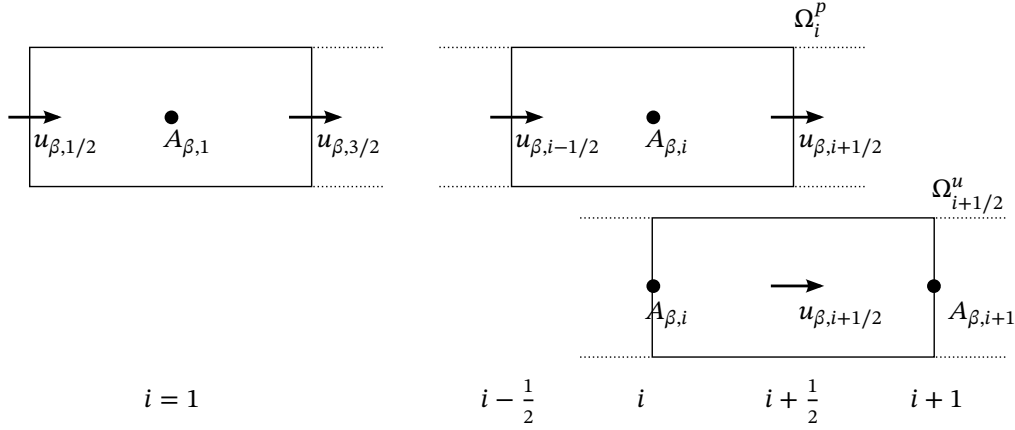 
\caption{Staggered grid layout including left boundary.}
\label{fig:GridLayout}
\end{figure}
\endgroup

%The semi-discrete mass and momentum equation for the gas phase read
%\begin{equation}
%\frac{\rd}{\rd t}\left( \left(A_{g} \rho_{g} \right)^{i} \right)+ \frac{(\rho_{g}A_{g})^{i+\frac{1}{2}}u_{g}^{i+\frac{1}{2}}-(\rho_{g}A_{g})^{i-\frac{1}{2}}u_{g}^{i-\frac{1}{2}}}{(\Delta x^{p})^{i}} = 0,
%\end{equation}
%\begin{multline}
%\frac{\rd }{\rd t}\left( \left( A_{g}\rho_{g}u_{g} \right)^{i+\frac{1}{2}} \right) + \frac{(\rho_{g}A_{g}u_{g}^{2})^{i+1}-(\rho_{g}A_{g}u_{g}^{2})^{i}}{(\Delta x^{u})^{i+\frac{1}{2}}} = \\
%-\frac{A_{g}^{i+\frac{1}{2}}(p^{i+1}-p^{i})}{(\Delta x^{u})^{i+\frac{1}{2}}} - \frac{\rho_{g}^{i+\frac{1}{2}}A_{g}^{i+\frac{1}{2}}g_{n}^{i+1/2} (h^{i+1}-h^{i})}{(\Delta x^{u})^{i+\frac{1}{2}}} - \rho_{g}^{i+1/2} A_{g}^{i+1/2} g_{s}^{i+1/2}.
%\end{multline}
%Expressions for the liquid phase are similar. The system is closed with the volume constraint, 
%\begin{equation}
%A_{g}^{i} + A_{l}^{i} = A^{i},
%\end{equation}
%\begin{multline}
%\frac{\rd }{\rd t}\left( \left( A_{l}\rho_{l}u_{l} \right)^{i+\frac{1}{2}} \right) + \frac{(\rho_{l}A_{l}u_{l}^{2})^{i+1}-(\rho_{l}A_{l}u_{l}^{2})^{i}}{(\Delta x^{u})^{i+\frac{1}{2}}} = \\ 
%-\frac{A_{l}^{i+\frac{1}{2}}(p^{i+1}-p^{i})}{(\Delta x^{u})^{i+\frac{1}{2}}} - \frac{\rho_{l}^{i+\frac{1}{2}}A_{l}^{i+\frac{1}{2}}g_{n}^{i+1/2} (h^{i+1}-h^{i})}{(\Delta x^{u})^{i+\frac{1}{2}}} - \rho_{l}^{i+1/2} A_{l}^{i+1/2} g_{s}^{i+1/2}.
%\end{multline}

We start with conservation of mass for the gas phase. Integration of the first equation in \eqref{eqn:PFTFM} in $s$-direction over a pressure volume gives:
\begin{equation}
\frac{\rd}{\rd t}\left(U_{1,i} \right) + \frac{U_{3,i+1/2}}{\Omega_{u,i+1/2}} - \frac{U_{3,i-1/2}}{\Omega_{u,i-1/2}} = 0.\label{eqn:mass_semidiscrete}
\end{equation}
A crucial detail in this equation is that the convective fluxes are directly expressed in terms of the momentum $\ul{U}_{3}$. For all volumes, the discretization is written as
\begin{equation}
\frac{\rd \ul{U}_{1}}{\rd  t} = \ul{F}_{1} (\ul{U}) = - D_p \left( \Omega_{u}^{-1} \ul{U}_{3} \right),
\end{equation}
where $D_p$ is a spatial differencing matrix, see \cite{Sanderse2018}. For clarity: $\ul{F}_{1}$ denotes the first component of the spatial discretization of $\vt{A} \dd{\vt{f}}{s} + \vt{A} \vt{S} + \vt{c} \dot{Q}$.  %and $\Omega_{p}$ a diagonal matrix containing the pressure-volume sizes $\Delta s_{i}$.

For conservation of momentum we proceed in a similar way. Integration of the third equation in \eqref{eqn:PFTFM} in $s$-direction over a velocity volume gives:
\begin{multline}\label{eqn:momentum_semidiscrete}
\frac{\rd}{\rd t}\left(U_{3,i+1/2} \right) + A_{33,i+1/2} (f_{3,i+1} - f_{3,i}) + A_{34,i+1/2} (f_{4,i+1} - f_{4,i}) + \\
  A_{33,i+1/2} S_{3,i+1/2} + A_{34,i+1/2} S_{4,i+1/2} - \left(\frac{A_{g} \rho_{l} \rho_{g}}{\hat{\rho}} \right)_{i+1/2} \Delta s_{i+1/2} \dot{Q}(t) = 0,
\end{multline}
where the notation $A_{ij}$ indicates an entry in the $\vt{A}$ matrix. \R{com14}\hl{Note that this discretization would also have been obtained when first discretizing the TFM and then applying the pressure-free transformation to it.}

%\begin{multline}
%\frac{\rd}{\rd t}\left(U_{3,i+1/2} \Delta s_{i+1/2} \right) + \left(\rho_{\beta} A_{\beta}\right)_{i+1} (u_{\beta,i+1})^2 - \left(\rho_{\beta} A_{\beta} \right)_{i} (u_{\beta,i})^2 = -A_{\beta,i+1/2}\left(p_{i+1} - p_{i} \right) \\ + \text{LG}_{\beta,i+1/2} + S_{\beta,i+1/2} \Delta s_{i+1/2}, \label{eqn:momentum_semidiscrete} %- \rho_{\beta,i+1/2} A_{\beta,i+1/2} \Delta s_{i+1/2} g_s  - \sum _{\substack{\gamma \in \{\text{L},\text{G},\text{W} \}\\\gamma \neq \beta }}  \tau_{\beta \gamma,i+1/2} P_{\beta \gamma,i+1/2} \Delta s_{i+1/2} + A_{\beta,i+1/2} F_{\text{body}} \Delta s_{i+1/2}, \label{eqn:momentum_semidiscrete} 
%\end{multline}
%and the level gradient terms for the gas and liquid are given by (+ for gas, - for liquid)
%\begin{equation}
%\text{LG}_{\beta,i+1/2} = \rho_{\beta} g_{n} \left( \left(h A_{\beta} \pm \frac{1}{12} P_{gl}^3 \right)_{i+1} - \left(h A_{\beta} \pm \frac{1}{12} P_{gl}^3 \right)_{i} \right). 
%\end{equation}
%The convective term in the momentum equation requires approximation; in the test cases in this work we have simply used a central approximation, $u_{\beta,i}=\frac{1}{2}(u_{\beta,i+1/2} + u_{\beta,i-1/2})$.
\hl{In the test cases in this article, either upwind (first-order accurate) or central (second-order accurate) schemes are used to express $f$ in terms of the conservative variables. In general, the scheme reads
\begin{equation}
f_{3,i} = (\rho_{g} A_{g})_{i} u_{g,i}^2 + K_{g,i}. % = \frac{U_{1,i}}{\Delta s_{i}} \left(\frac{1}{2} \left(\frac{U_{3,i+1/2}}{\Delta s_{i+1/2}} + \frac{U_{3,i+3/2}}{\Delta s_{i+3/2}} \right) \right)^2.
\end{equation}
In the central scheme we have $u_{g,i}=\frac{1}{2}(u_{g,i-1/2} + u_{g,i+1/2})$, whereas in the first-order upwind scheme $u_{g,i}=u_{g,i-1/2}$ if $u_{g,i-1/2}>0$. $A_{g}$ and $u_{g}$ follow from equation \eqref{eqn:primitive_vars}, and $K_{g}$ from equation \eqref{eqn:compressible_HG}. A similar expression holds for $f_{4}$, with subscripts $g$ replaced by $l$. For more details, we refer to \cite{Sanderse2018}. The particular choice for $f_{3}$ and $f_{4}$ does not affect the derivation of the discrete pressure-free model. If desired, higher-order methods for approximating $f_{3}$ and $f_{4}$ can be easily employed instead.}

In summary, the discretization of the momentum equation of the gas phase is written as
\begin{equation}
\frac{\rd \ul{U}_{3}}{\rd  t} = \ul{F}_{3}(\ul{U}),
\end{equation}
and similar for $\ul{U}_{4}$. We stress that in contrast to the original TFM, $\ul{F}_{3}$ does not contain the pressure, but instead contains the $\vt{A}$-matrix and volumetric source term involving $\dot{Q}$. The entire spatial discretization (mass and momentum equations) can be summarized as
\begin{equation}\label{eqn:ODE}
\frac{\rd \ul{U}(t)}{\rd  t} = \ul{F}(\ul{U}(t),t).
\end{equation} 
%where the matrices $\Omega_p$ and $\Omega_u$ have been absorbed in the definition of $\ul{F}$.

We will now check whether the volume and volumetric flow constraints are satisfied with this spatial discretization. The first-order time-derivative of the volume constraint is given by
\begin{equation}
\begin{split}
\frac{\rd }{\rd t} \left( \ul{A}_{g} + \ul{A}_{l} \right) &= \frac{\rd }{\rd t} \left( \frac{\Omega_{p}^{-1} \ul{U}_{1}}{\rho_{g}} + \frac{\Omega_{p}^{-1} \ul{U}_{2}}{\rho_{l}} \right) \\
&= \Omega_{p}^{-1} \frac{\rd }{\rd t} \left( \frac{ \ul{U}_{1}}{\rho_{g}} + \frac{\ul{U}_{2}}{\rho_{l}} \right) \\
&= - \Omega_{p}^{-1} \left( \frac{ \ul{F}_{1}}{\rho_{g}} + \frac{\ul{F}_{2}}{\rho_{l}} \right) \\
&= - \Omega_{p}^{-1} D_{p} \left( \frac{\Omega_{u}^{-1} \ul{U}_{3}}{\rho_{g}} + \frac{\Omega_{u}^{-1}  \ul{U}_{4}}{\rho_{l}} \right).
\end{split}
\end{equation}
The second-order time-derivative then follows as  
\begin{equation}\label{eqn:d2A}
\begin{split}
\frac{\rd^2 }{\rd t^2} \left( \ul{A}_{g} + \ul{A}_{l} \right) &= - \frac{\rd }{\rd t} \left(\Omega_{p}^{-1} D_{p} \left( \frac{\Omega_{u}^{-1} \ul{U}_{3}}{\rho_{g}} + \frac{\Omega_{u}^{-1} \ul{U}_{4}}{\rho_{l}} \right) \right), \\
&= - \Omega_{p}^{-1} D_{p} \Omega_{u}^{-1} \frac{\rd }{\rd t} \left( \frac{\ul{U}_{3}}{\rho_{g}} + \frac{\ul{U}_{4}}{\rho_{l}} \right), \\
&= - \Omega_{p}^{-1} D_{p} \Omega_{u}^{-1}  \left( \frac{ \ul{F}_{3}}{\rho_{g}} + \frac{\ul{F}_{4}}{\rho_{l}} \right).
\end{split}
\end{equation}
The last term can be simplified by realizing that the $\vt{A}$ matrix in the PF-TFM has been constructed such that:
\begin{equation}\label{eqn:F3F4}
\frac{\ul{F}_{3}}{\rho_{g}} + \frac{\ul{F}_{4}}{\rho_{l}} = \Omega_u \dot{\ul{Q}}(t),
\end{equation}
where $\dot{\ul{Q}} \in \mathbb{R}^{N_u} = \ul{1} \dot{Q}(t)$ is a vector with $Q(t)$ in each entry. The correctness of this expression can be easily checked by substituting the expressions for $\ul{F}_{3}$ and $\ul{F}_{4}$ (see equation \eqref{eqn:momentum_semidiscrete}), and for $\vt{A}$ and $\vt{c}$, see  equation \eqref{eqn:expressionAC}. See also equation \eqref{eqn:F3}. Consequently, equation \eqref{eqn:d2A} gives
\begin{equation}\label{eqn:d2A2}
\begin{split}
\frac{\rd^2 }{\rd t^2} \left( \ul{A}_{g} + \ul{A}_{l} \right) &= - \Omega_{p}^{-1} D_{p} \dot{\ul{Q}}(t), \\
& = 0.
\end{split}
\end{equation}
The last equality holds because application of the differencing matrix $D_p$ to a constant vector yields 0.

Integrating equation \eqref{eqn:d2A2} twice in time, given an initial condition that satisfies the volume constraint ($\ul{A}_{g}(t_{0}) + \ul{A}_{l} (t_0) = A$), then leads to
\begin{equation}
\ul{A}_{g}(t) + \ul{A_{l}}(t) = A.
\end{equation}
In other words, the volume constraint remains satisfied in time by employing a spatial discretization on a staggered grid.

%The system is closed with the volume constraint \eqref{eqn:volume_constraint}, which is written in terms of the phase masses $m_{\beta}$  as: 
%\begin{equation}
%\frac{\ul{U}_{1}}{\rho_{g}} + \frac{\ul{U}_{2}}{\rho_{l}} - A = 0.
%\end{equation}

% This ensures a discrete coupling between the mass and momentum equations in the same way as in the continuous case, where the incompressible pressure equation was derived by equating the time differentiation of the flux terms in the mass equations to the spatial differentiation of the unsteady terms in the momentum equations. 

%In order to derive the approximation of the $\vt{H}$ matrix, we perform on the discrete level the same steps as on the continuous level. First write:
%\begin{multline}
%\frac{\rd}{\rd t}\left(U_{3,i+1/2} \Delta s_{i+1/2} \right) + A_{33,i+1/2} (f_{3,i+1} - f_{3,i-1}) + A_{34,i+1/2} (f_{4,i+1} - f_{4,i-1}) + \\
%  A_{33,i+1/2} S_{3,i+1/2} + A_{34,i+1/2} S_{4,i+1/2} = 0.
%\end{multline}

%\begin{equation}
%\ul{F}(U) = 
%\end{equation}

\section{Explicit time discretization: efficient and constraint-consistent}\label{sec:time_discr}
\subsection{Basic formulation}
Whereas the previous section showed that the pressure-free TFM posed requirements on the spatial discretization in order to be constraint-consistent, no further requirements are necessary for the temporal discretization: a simple forward Euler time discretization already keeps the constraint property. The extension to generic explicit Runge-Kutta methods is then straightforward. Note that due to the absence of the pressure, the spatially discretized system can be advanced straightforwardly in time (without the need to solve a Poisson equation). 

The forward Euler step is given by
\begin{equation}\label{eqn:FE}
\ul{U}^{n+1} = \ul{U}^{n}  +  \Delta t \ul{F}(\ul{U}^{n}).
\end{equation}
An important term present in $\ul{F}(\ul{U}^{n})$ is the discrete approximation to the last term in equation \eqref{eqn:momentum_semidiscrete}, being the volumetric flow source term. We split $\ul{F}(\ul{U})$ as 
\begin{equation}
\ul{F}(\ul{U}) = \ul{\hat{F}}(\ul{U}) + \ul{c}(\ul{U}) \dot{Q}(t),
\end{equation}
The forward Euler discretization of the last term, as given by equation \eqref{eqn:FE}, yields:
\begin{equation}\label{eqn:optionA}
\text{option A:} \qquad \qquad \Delta t \ul{c}(\ul{U}^{n}) \dot{Q}(t^{n}).
\end{equation}
An alternative formulation (option B) of the source term will be discussed in section \ref{sec:high_order}.

%Note that, with some abuse of notation, the term in $\ul{F}(\ul{U}^{n})$ which involves $\dot{Q}$ is not discretized as $\dot{Q}(t^{n})$ but as $(Q(t^{n+1}) - Q(t^{n}))/\Delta t$ (this requires less differentiability of $Q(t)$ and will be used to keep the volumetric flow balance exact).

%where $\ul{F}(U)$ is given by
%\begin{equation}
%\ul{F}(U) = - 
%\begin{pmatrix}
%f_{1} (U_{3}) \\
%f_{2} (U_{4}) \\
%f_{3} (U) \\
%f_{4} (U)
%\end{pmatrix}
%- 
%\begin{pmatrix}
%0 \\
%0 \\
%S_{3} \\
%S_{4}
%\end{pmatrix}
%+
%\begin{pmatrix}
%0 \\ 
%0 \\
%\frac{A_{g} \rho_{l} \rho_{g}}{\hat{\rho}} \\
%\frac{A_{l} \rho_{l} \rho_{g}}{\hat{\rho}}
%\end{pmatrix}
%\dot{Q}(t).
%\end{equation}

As before, the key question is whether the constraint is satisfied at the new time level $t^{n+1}$ (assuming that it was satisfied at the current time level $t^{n}$):
\begin{equation}
\begin{split}
\ul{A}_{g}^{n+1} + \ul{A}_{l}^{n+1} - A &= \frac{\Omega_{p}^{-1} \ul{U}_{1}^{n+1}}{\rho_{g}} + \frac{\Omega_{p}^{-1} \ul{U}_{2}^{n+1}}{\rho_{l}} - A \\
&= \Omega_{p}^{-1} \left(\frac{\ul{U}_{1}^{n} + \Delta t  \ul{F}_{1} (\ul{U}^{n}) }{\rho_{g}} + \frac{\ul{U}_{2}^{n} + \Delta t  \ul{F}_{2} (\ul{U}^{n}) }{\rho_{l}} \right) - A\\
&= \underbrace{\ul{A}_{g}^{n} + \ul{A}_{l}^{n} - A}_{0} - \Delta t \Omega_{p}^{-1} D_p \left( \frac{\Omega_{u}^{-1} \ul{U}_{3}^{n}}{\rho_{g}} + \frac{\Omega_{u}^{-1}  \ul{U}_{4}^{n}}{\rho_{l}} \right) \\
&= - \Delta t \Omega_{p}^{-1} D_p \Omega_{u}^{-1} \left( \frac{ \ul{U}_{3}^{n}}{\rho_{g}} + \frac{ \ul{U}_{4}^{n}}{\rho_{l}} \right). \label{eqn:constraint_fullydiscrete}
\end{split}
\end{equation}
In order to simplify this equation further, we insert the expression for the momentum equation and employ property \eqref{eqn:F3F4}
\begin{equation}\label{eqn:Q}
\begin{split}
 \frac{\ul{U}_{3}^{n}}{\rho_{g}} + \frac{\ul{U}_{4}^{n}}{\rho_{l}} &=  \frac{\ul{U}_{3}^{n-1} + \Delta t \ul{F}_{3} (\ul{U}^{n-1})}{\rho_{g}}  + \frac{\ul{U}_{4}^{n-1}+ \Delta t \ul{F}_{4} (\ul{U}^{n-1})}{\rho_{l}} \\
 &=  \frac{\ul{U}_{3}^{n-1}}{\rho_{g}}  + \frac{\ul{U}_{4}^{n-1}}{\rho_{l}}  + \Omega_{u} \dot{\ul{Q}}(t^{n-1}). %\Omega_{u} \left( \ul{Q}(t^{n}) - \ul{Q}(t^{n-1}) \right).
\end{split}
\end{equation}
Here we used option A for the discretization of the volumetric source term. Given an initial condition for $\ul{U}_{3}$ and $\ul{U}_{4}$ at $t=t_{0}$ which satisfies the volumetric flow constraint,
\begin{equation}
D_{p} \Omega_{u}^{-1} \left( \frac{\ul{U}_{3}^{1}}{\rho_{g}} + \frac{\ul{U}_{4}^{1}}{\rho_{l}} \right) = 0, % \ul{Q}(t_0),
\end{equation}
and given that $D_{p} \dot{\ul{Q}}(t)=0$, the volumetric flow constraint can be written as a telescopic sum over all previous time steps:
\begin{equation}\label{eqn:diffQ}
D_p \Omega_{u}^{-1} \left( \frac{\ul{U}_{3}^{n}}{\rho_{g}} + \frac{\ul{U}_{4}^{n}}{\rho_{l}} \right) = \ldots = D_p \Omega_{u}^{-1} \left( \frac{\ul{U}_{3}^{1}}{\rho_{g}} + \frac{\ul{U}_{4}^{1}}{\rho_{l}} \right) =0.
\end{equation}
The volume constraint at the new time level, equation \eqref{eqn:constraint_fullydiscrete}, can thus be evaluated as
\begin{equation}
\ul{A}_{g}^{n+1} + \ul{A}_{l}^{n+1} - A = 0.
\end{equation}
In summary, when applying the forward Euler method to equation \eqref{eqn:ODE} and starting from consistent initial conditions, both the volume constraint and the volumetric flow constraint are satisfied at each time step.

Note that neither equation \eqref{eqn:Q}, nor equation \eqref{eqn:diffQ}, guarantees that the solution $\ul{U}^{n+1}$ will satisfy the exact flow rate $Q(t^{n+1})$, since only the time-derivative $\dot{Q}(t^{n})$ has been used. However, when instead of option A, we apply the following first-order discretization of the volumetric flow source term
\begin{equation}\label{eqn:optionB}
\text{option B:} \qquad \qquad \ul{c}(\ul{U}^{n}) (Q(t^{n+1}) - Q(t^{n}))
\end{equation}
into equation \eqref{eqn:Q}, not only the volumetric flow constraint is satisfied, but in addition the actual value of the volumetric flow stays equal to the specified volumetric flow when marching in time:
\begin{equation}
\begin{split}
 \frac{\ul{U}_{3}^{n}}{\rho_{g}} + \frac{\ul{U}_{4}^{n}}{\rho_{l}} = \Omega_{u} \ul{Q}(t^{n}).
\end{split}
\end{equation}
In other words, when adapting the forward Euler discretization for the volumetric flow source term, it is possible to satisfy the volumetric flow constraint \textit{and} the actual value of $Q(t)$ exactly. The generalization of this idea to high-order time integration methods requires careful considerations, as will be discussed next.

\subsection{High-order accuracy and integration of $\dot{Q}(t)$}\label{sec:high_order}
%Consequently we have
%\begin{equation}\label{eqn:diffQ}
%D_p \Omega_{u}^{-1} \left( \frac{\ul{U}_{3}^{n}}{\rho_{g}} + \frac{\ul{U}_{4}^{n}}{\rho_{l}} \right) = 0.
%%D_p \Omega_{u}^{-1} \left( \frac{\ul{U}_{3}^{n-1}}{\rho_{g}} + \frac{\ul{U}_{4}^{n-1}}{\rho_{l}} \right) = \ldots = D_p \Omega_{u}^{-1} \left( \frac{\ul{U}_{3}^{1}}{\rho_{g}} + \frac{\ul{U}_{4}^{1}}{\rho_{l}} \right) =  0.
%\end{equation}
%Note that this derivation hinges on the use of $(Q(t^{n+1}) - Q(t^{n}))/\Delta t$ instead of $\dot{Q}(t)$ in equation \eqref{eqn:FE}. 
The high-order time integration methods we consider are explicit Runge-Kutta (RK) methods. Explicit RK methods form an excellent combination of accuracy and stability for convection-dominated systems and, in contrast to the standard TFM, they can be applied to the proposed PF-TFM in a straightforward manner, because the constraint has been eliminated. When considering option A for handling the volumetric flow source term, the resulting method reads:
%\begin{align}
\begin{empheq}[left=\text{option A}\quad,innerbox=\fbox]{align}
\ul{U}^{n,i} &= \ul{U}^{n}  + \Delta t \sum_{j=1}^{i-1} a_{ij} \ul{F} (\ul{U}^{n,j},t^{j}), \label{eqn:ERK_stages}\\ 
\ul{U}^{n+1} &= \ul{U}^{n}  + \Delta t \sum_{i=1}^{s} b_{i} \ul{F} (\ul{U}^{n,i},t^{i}), \label{eqn:ERK_step}
\end{empheq}
%\end{align}
where $a$ and $b$ are the coefficients of the Butcher tableau, and $s$ the number of stages. Since explicit RK methods can be seen as a combination of forward Euler steps, it is straightforward to prove that they satisfy the two constraints, both for the stage values $\ul{U}^{n,i}$ and for $\ul{U}^{n+1}$. 

When considering option B for handling the volumetric source term, there is an open question at which time level to evaluate $\ul{c}(\ul{U})$ at the intermediate stages. One possibility that leads to an exact volumetric flow is to take at each stage 
\begin{equation}
\ul{c}(\ul{U}^{n}) (Q(t^{n,i}) - Q(t^{n})).
\end{equation}
However, this \R{com6} choice does not fit in the Runge-Kutta framework, and the resulting method suffers from loss of order of accuracy. 
% Our proposal is to keep this property by extending equation \eqref{eqn:optionB} to a generic Runge-Kutta method

The question how to employ option B in conjunction with a high-order method can be answered with the following insight. \textit{The time discretization corresponding to option B, equation \eqref{eqn:optionB}, is obtained when deriving the pressure-free model on the fully discrete level instead of on the continuous level.} In other words: option A is obtained by first eliminating the pressure from the two-fluid model, and then discretizing the resulting system; option B is obtained by first discretizing the two-fluid model, and then eliminating the pressure. The details of this derivation are shown in \ref{sec:PF_fullydiscrete}. With this insight, a high-order RK method for option B is straightforward to construct: eliminate the pressure from the (fully discrete) high-order RK methods proposed in \cite{Sanderse2018}, and rewrite as a time integration method for the PF-TFM. 

We focus on the case $s=3$, i.e.\ a three-stage third-order method, for which the following method results:
%\begin{align}
\begin{empheq}[left=\text{option B}\quad,innerbox=\fbox]{align}
\ul{U}^{n,1} &= \ul{U}^{n}, \label{eqn:RK_optionB_1}\\
\ul{U}^{n,2} &= \ul{U}^{n}  + \Delta t \sum_{j=1}^{1} \left[ a_{2j} \ul{\hat{F}} (\ul{U}^{n,j},t^{j}) \right] + \ul{c}(\ul{U}^{n,1}) (Q^{n,2} - Q^{n}), \label{eqn:RK_optionB_2}\\ 
\begin{split}
\ul{U}^{n,3} &= \ul{U}^{n}  + \Delta t \sum_{j=1}^{2} \left[ a_{3j} \ul{\hat{F}} (\ul{U}^{n,j},t^{j})  \right]  + \ul{c}(\ul{U}^{n,1}) \frac{a_{31}}{a_{21}} (Q^{n,2} - Q^{n})  + \\
& \quad \ul{c}(\ul{U}^{n,2}) \left( (Q^{n,3} - Q^{n})  - \frac{a_{31}}{a_{21}} (Q^{n,2} - Q^{n}) \right),
\end{split} \label{eqn:RK_optionB_3} \\
\begin{split}
\ul{U}^{n+1} &= \ul{U}^{n}  + \Delta t \sum_{i=1}^{3}\left[ b_{i} \ul{\hat{F}} (\ul{U}^{n,i},t^{i})  \right] + \ul{c}(\ul{U}^{n,1}) \frac{b_{1}}{a_{21}} (Q^{n,2} - Q^{n})  + \\ 
&  \quad \ul{c}(\ul{U}^{n,2}) \left( \frac{b_{2}}{a_{32}}(Q^{n,3} - Q^{n})  - \frac{b_{2}}{a_{21}} \frac{a_{31}}{a_{32}} (Q^{n,2} - Q^{n}) \right)  + \\
&  \quad \ul{c}(\ul{U}^{n,3}) \left( (Q^{n+1} - Q^{n}) -  \frac{b_{2}}{a_{32}}(Q^{n,3} - Q^{n})  -  \left( \frac{b_{1}}{a_{21}} - \frac{b_{2}}{a_{21}} \frac{a_{31}}{a_{32}}\right) (Q^{n,2} - Q^{n}) \right).
\end{split} \label{eqn:RK_optionB_4}
\end{empheq}
%\end{align}
%\begin{align}
%\ul{U}^{n,i} &= \ul{U}^{n}  + \Delta t \sum_{j=1}^{i-1} a_{ij} \ul{\hat{F}} (\ul{U}^{n,j},t^{j})  + \sum_{j=1}^{i} \tilde{a}_{ij} \ul{c} (\ul{U}^{n,j}) (Q^{n,j}-Q^{n})\\ 
%\ul{U}^{n+1} &= \ul{U}^{n}  + \Delta t \sum_{i=1}^{s} b_{i} \ul{F} (\ul{U}^{n,i},t^{i}).
%\end{align}
%
%For a three stage ($s=3$) method, the tableau $\tilde{a}$ is given by
%\begin{equation}\label{eqn:RK3_tableaux}
%\def\arraystretch{1.25}
%\begin{array}{ccc}
%\multicolumn{3}{c}{\text{}}\\
% 0 \\
% \tilde{a}_{21}  \\
%  \tilde{a}_{31}  & \tilde{a}_{32} \\
%\hline 
%\tilde{b}_{1} & \tilde{b}_{2} & \tilde{b}_{3}
%\end{array}
%=
%\begin{array}{ccc}
%\multicolumn{3}{c}{\text{}}\\
% 0 \\
%1  \\
%  \frac{a_{31}}{a_{21}}  & \tilde{a}_{32} \\
%\hline 
%\tilde{b}_{1} & \tilde{b}_{2} & \tilde{b}_{3}
%\end{array}
%\end{equation}
It can be checked through substitution that the volumetric flow remains equal to the specified volumetric flow $Q(t)$, both at the stage levels and at the new time level. It is important to remark that, although this is strictly speaking not a Runge-Kutta method applied to \eqref{eqn:ODE}, it \textit{is} a Runge-Kutta method applied to the original TFM. The order of accuracy of this method was analyzed in \cite{Sanderse2018}, and a specific third-order method was proposed:
\begin{equation}
\def\arraystretch{1.5}
\begin{array}{c|c}
c & a \\
\hline 
 & b
\end{array}
\quad = \quad
\begin{array}{c|ccc}
%\multicolumn{4}{c}{\text{}}\\
0& 0 \\
\frac{1}{2} & \frac{1}{2}  \\
1 & -1 & 2 \\
\hline 
& \frac{1}{6} & \frac{2}{3} & \frac{1}{6}\\
%\multicolumn{4}{c}{\text{}}\\
%\multicolumn{4}{c}{\text{RK3 in \cite{Sanderse2018}}}\\
\end{array}
\end{equation}
This method is such that the lower diagonal entries ($a_{21}$, $a_{32}$) are nonzero (this is required in equations \eqref{eqn:RK_optionB_3}-\eqref{eqn:RK_optionB_4}), and order reduction that can result from unsteady boundary conditions is avoided (details in next section). The $c$-coefficients are used to determine the intermediate time levels ($t^{i} = t^{n}+c_{i} \Delta t$) and should not be confused with the vector $\vt{c}$ or $\ul{c}$ as present in the PF-TFM.

\R{com14_2}\hl{Note that the temporal accuracy (third order) of this method is higher than of the spatial discretization (second order for the central scheme, and first order for the upwind scheme). This is because first- and second-order Runge-Kutta methods are not stable for pure convection problems (discretized with central schemes), independent of the time step, in contrast to third-order methods. Furthermore, the third-order method leaves room for future improvements in terms of high-order spatial discretization methods.}

\R{com2}\hl{Although equations \eqref{eqn:RK_optionB_1} - \eqref{eqn:RK_optionB_4} might seem cumbersome from an implementation point of view, one can use an existing Runge-Kutta implementation and simply add the additional vectors on the right hand-side that involve $\ul{c}$ and $Q$. The effective coefficients of the scheme, which involve combinations of Runge-Kutta coefficients, can be easily precomputed.}

%In summary, time integration with an explicit RK method preserves the volume constraint and the volumetric flow constraint. 
In summary, we have proposed high-order time integration of the PF-TFM for both `weak' (option A) and `strong' (option B) imposition of the volumetric flow source term. In both cases the two constraints are satisfied, and in the latter case also the volumetric flow remains equal to the specified value. The second approach requires a specialized Runge-Kutta method, which is slightly more involved to implement. Note that in the special case that $\dot{Q}(t)=0$, we have $\hat{\ul{F}}=\ul{F}$ and option B becomes a classic Runge-Kutta method, equivalent to option A. In this case we will resort to a classic four-stage fourth order Runge-Kutta method.
%In addition, it is possible (option B) to exactly track the volumetric flow $Q(t)$ by adapting the treatment of the volumetric flow source term in the RK method. 

%However, this comes at the price of a formal loss in order of accuracy of the time integration method. We therefore favour the use of option A: integration of $\dot{Q}(t)$ in time with the Runge-Kutta method. The associated error in the actual volumetric flow is $\mathcal{O}(\Delta t^4)$ for the fourth order Runge-Kutta method that we employ. 
%Note that a `standard' incompressible TFM discretization would also not satisfy

\subsection{Boundary conditions}
The boundary condition treatment follows the characteristic approach outlined in \cite{Sanderse2018}, which has the advantage that it does not require the pressure in the formulation. Here we shortly summarize the (important) case of unsteady inflow conditions, with prescribed gas and liquid (mass) flows, $I_{g}(t)$ and $I_{l}(t)$, respectively. There are two choices for the boundary conditions: prescription in strong form (prescribe the actual value of $I_{g}$ and $I_{l}$ at each time step), or in weak form (prescribe the time-derivatives $\dot{I}_{g}$ and $\dot{I}_{l}$ and integrate in time with the Runge-Kutta method).

It is important to note that \textit{the boundary condition prescription should be consistent with the discretization of the volumetric source term} $\ul{c}(\ul{U}) \dot{Q}(t)$ in order to make sure that $\ul{Q}$ remains uniform in space, so that the volumetric flow constraint is satisfied. The weak boundary condition prescription is consistent with option A. The strong boundary condition prescription is consistent with option B. 
%Since we have chosen option A for reasons outline above, the boundary conditions will be prescribed in weak form.

\subsection{Keeping the constraint exact in time by preventing machine error accumulation}
In deriving equation \eqref{eqn:constraint_fullydiscrete} the error in the volume constraint at the previous time step was imposed to be exactly equal to zero. In actual computations, the value of $A_{g}^{n} + A_{l}^{n} - A$ might not exactly equal zero due to the presence of machine precision errors. By simply leaving this term in the numerical algorithm, one can avoid accumulation of machine precision errors that could spoil the accuracy of the computations \cite{Hirt1967a,Veldman2013b}. In other words, one should \textit{not} impose the volume constraint term in \eqref{eqn:constraint_fullydiscrete} to be zero, but instead simply compute its value based on the solution at the last time step ($A_{g}^{n} + A_{l}^{n} - A$) and use this while time stepping. In practice we achieve this by computing the error in the volume constraint after each stage of the Runge-Kutta method:
\begin{align}
\ul{\varepsilon}_A^{n,i} &= \Omega_{p}^{-1}  \left( \frac{\ul{U}_{1}^{n,i}}{\rho_g} + \frac{\ul{U}_{2}^{n,i}}{\rho_l} \right) - A,
%\ul{\varepsilon}_Q^{n,i} &= \Omega_{u}^{-1} \left( \frac{ \ul{U}_{3}^{n,i}}{\rho_g} + \frac{\ul{U}_{4}^{n,i}}{\rho_l} \right)  - \ul{Q}(t^{i}),
\end{align}
and adding this term to the solution after each stage:
\begin{align}
\hat{\ul{U}}^{n,i} = \ul{U}^{n,i} + \ul{\varepsilon}^{n,i},
\end{align}
where
\begin{equation}
 \ul{\varepsilon}^{n,i} = -\frac{1}{2} \begin{pmatrix}
 \rho_{g} \Omega_{p} \ul{\varepsilon}_{A}^{n,i} \\
  \rho_{l} \Omega_{p} \ul{\varepsilon}_{A}^{n,i} \\
  0 \\ % \rho_{g} \Omega_{u} \ul{\varepsilon}_{Q}^{n,i} \\
0  %   \rho_{l} \Omega_{u} \ul{\varepsilon}_{Q}^{n,i}
 \end{pmatrix}.
\end{equation}
Note that this does not affect $\ul{U}_{3}$ and $\ul{U}_{4}$, so it does not change the volumetric flow.

\section{Results}\label{sec:results}

\subsection{Kelvin-Helmholtz instability}\label{sec:KH}
We perform the classic viscous Kelvin-Helmholtz instability test case, see e.g.\ \cite{Liao2008,Sanderse2018} and table \ref{tab:VKH} for parameter values.  We choose values for $A_{l}$ and $u_{l}$, $u_{l} = \SI{1}{m/s}$ and $\alpha_{l}=A_{l}/A = 0.9$ and then compute the gas velocity and the body force necessary to sustain the steady solution:
\begin{equation}
u_{g} = \SI{8.01}{m/s}, \qquad \frac{\rd p_{0}}{\rd s} = \SI{-87.87}{Pa/m}.
\end{equation}
The velocity difference $u_{g}-u_{l}$ is below the `Kelvin-Helmholtz' instability limit \cite{Liao2008,Sanderse2018}, which means that, at least initially, the initial boundary value problem is well-posed. In addition, the conditions are such that the solution is linearly \textit{unstable}, so we can study the growth of waves. We impose that $\dot{Q}=0$, so that option A and B are equivalent. In this case, we use a standard four-stage, fourth-order RK method to integrate in time.

\begin{table}[hbtp]
\centering
\caption{Parameter values for the test case with the Kelvin-Helmholtz instability. \label{tab:VKH}}
\begin{tabular}{lrl}
\toprule
parameter & value & unit \\
\midrule
$\rho_{l}$  & $1000$ & \si{kg/m^{3}} \\ 
$\rho_{g}$  & $1.1614$ & \si{kg/m^{3}} \\ 
$R$ & $0.039$ & \si{m} \\
$g$ & $9.8$ & \si{m/s^{2}} \\
$\mu_{g}$ & $1.8 \cdot 10^{-5}$ & \si{Pa.s} \\
$\mu_{l}$ & $8.9 \cdot 10^{-4}$ & \si{Pa.s} \\ 
$\epsilon$ & $10^{-8}$ & \si{m} \\
$L$ & $1$ & \si{m} \\
\bottomrule
\end{tabular}
\end{table}

We perturb the steady state by imposing a sinusoidal disturbance with wavenumber $k=2\pi$ (unit \si{1/m}) and a small amplitude. Linear stability analysis \cite{Liao2008,Sanderse2017} gives the following angular frequencies $\omega$ (unit \si{1/s}):
\begin{align}\label{eqn:omega_exact}
\omega_{1} = 3.22 + 2.00 i, \\ 
\omega_{2} =  10.26 - 1.61 i.
\end{align}
The negative imaginary part of $\omega_{2}$ indicates the instability in the solution. We choose the initial perturbation related to $\omega_{1}$ to be zero, implying that a single wave with frequency $\omega_{2}$ results, and the linearized solution is
\begin{equation}\label{eqn:analytical_linearized}
\vt{U} (s,t) = \vt{U}^{0} + \operatorname{Re} \left[ \Delta \vt{U} e^{i(\omega_2 t - ks)} \right],
\end{equation}
where  $\Delta \vt{U}$ is obtained by choosing the liquid hold-up fraction perturbation as $\hat{\alpha}_{l}=10^{-3}$, and then computing the perturbations in the gas and liquid velocity from the dispersion analysis \cite{Liao2008}. The initial condition which follows by taking $t=\SI{0}{s}$ does in general not satisfy \eqref{eqn:IC2} exactly, and we therefore perform a projection step to make the initial conditions consistent, see \cite{Sanderse2018}, in such a way that the volumetric flow rate stays exact. 

\begin{figure}[h!]
\centering
\includegraphics[width=\textwidth]{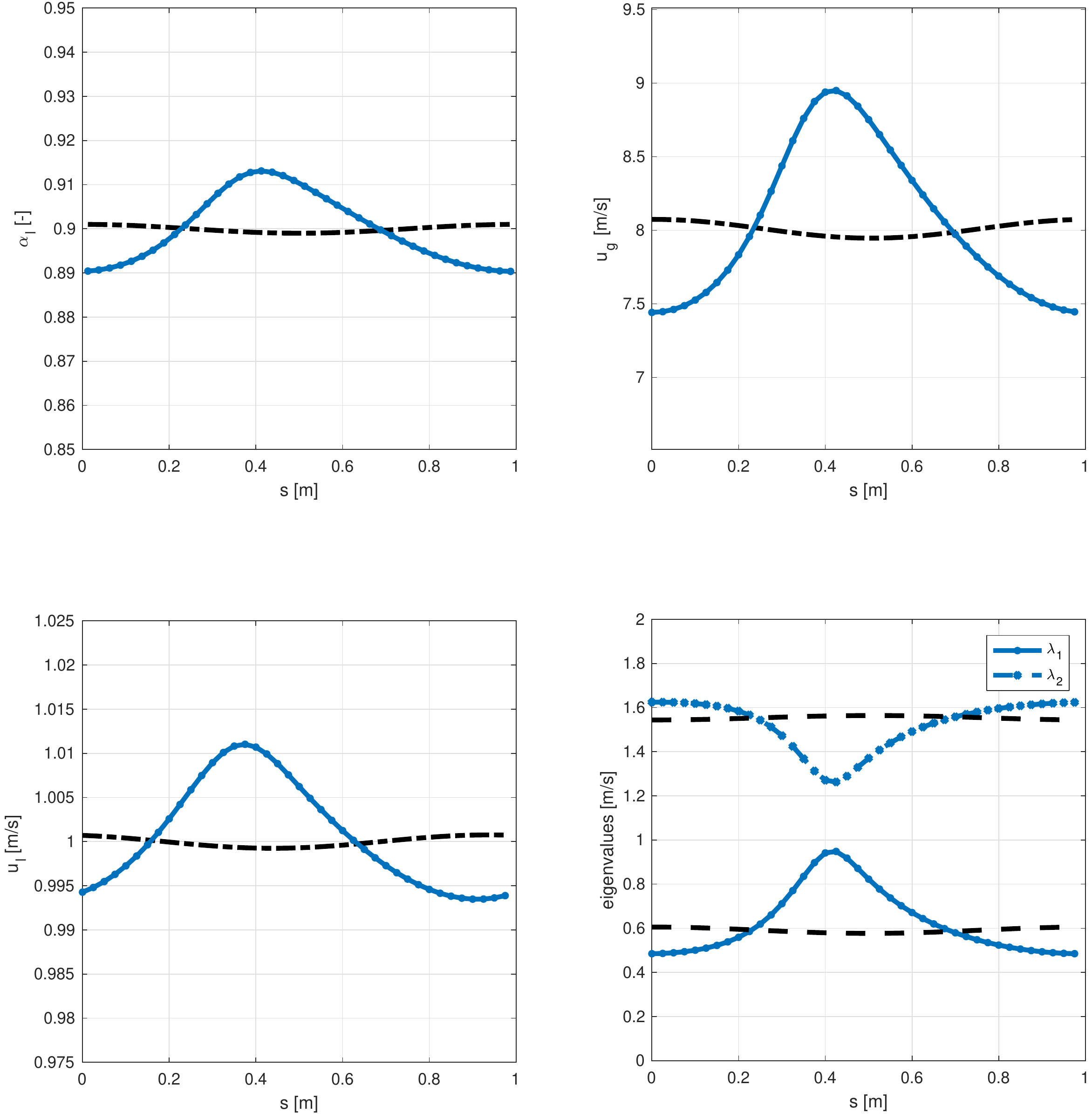}
\caption{Solution at $t=\SI{1.5}{s}$ with $N=40$ and $\Delta t = \SI{1/100}{s}$ (blue), including initial condition (black dashed).\label{fig:fullsolution_KH_INC_PFTFM_N_40_dt001}}
\end{figure}

\begin{figure}[h!]
\centering
\begin{subfigure}[b]{.49\textwidth}
	\centering
\includegraphics[width=\textwidth]{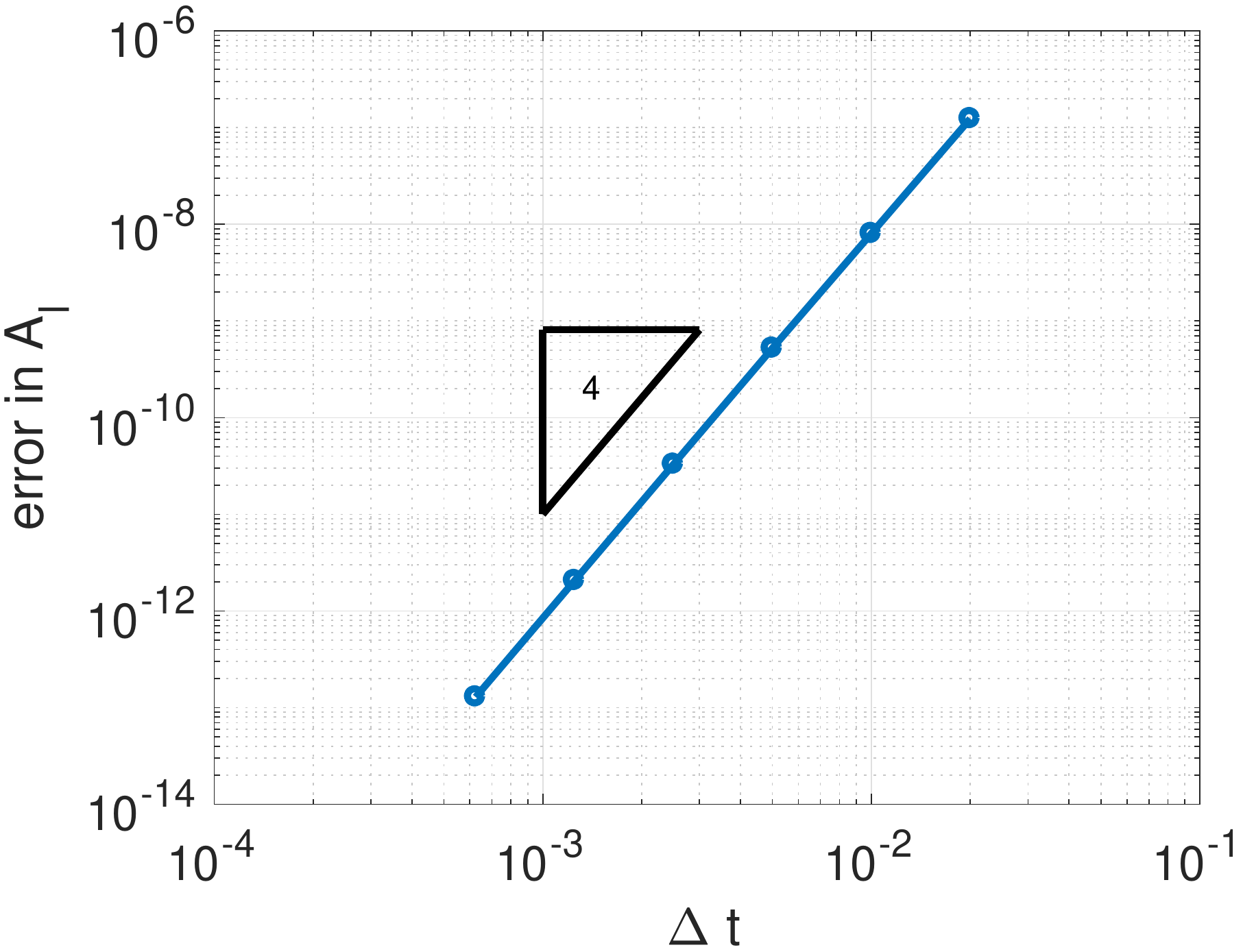}
\caption{Error in hold-up.\label{fig:convergence_KH_INC_PFTFM_N_40_A_l}}
	\end{subfigure}
	\hfill
\begin{subfigure}[b]{.49\textwidth}
	\centering
\includegraphics[width=\textwidth]{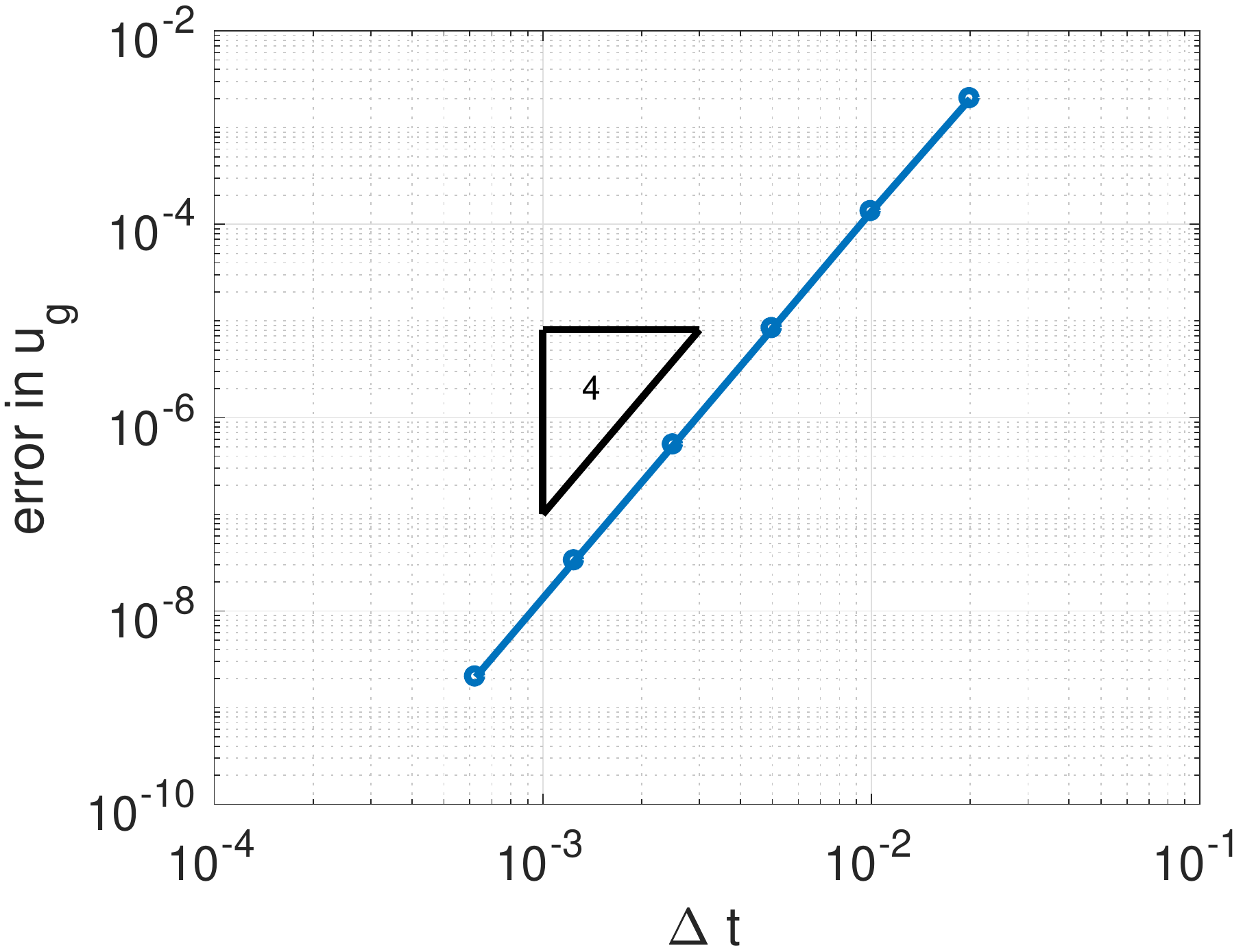}
\caption{Error in gas velocity. \label{fig:convergence_KH_INC_PFTFM_N_40_u_g}}
	\end{subfigure}
\caption{Fourth-order convergence of the error at $t=\SI{1.5}{s}$ for the Kelvin-Helmholtz test case.\label{fig:convergence_KH_INC_PFTFM_N_40}}
\end{figure}

\begin{figure}[h!]
\centering
\includegraphics[width=0.6\textwidth]{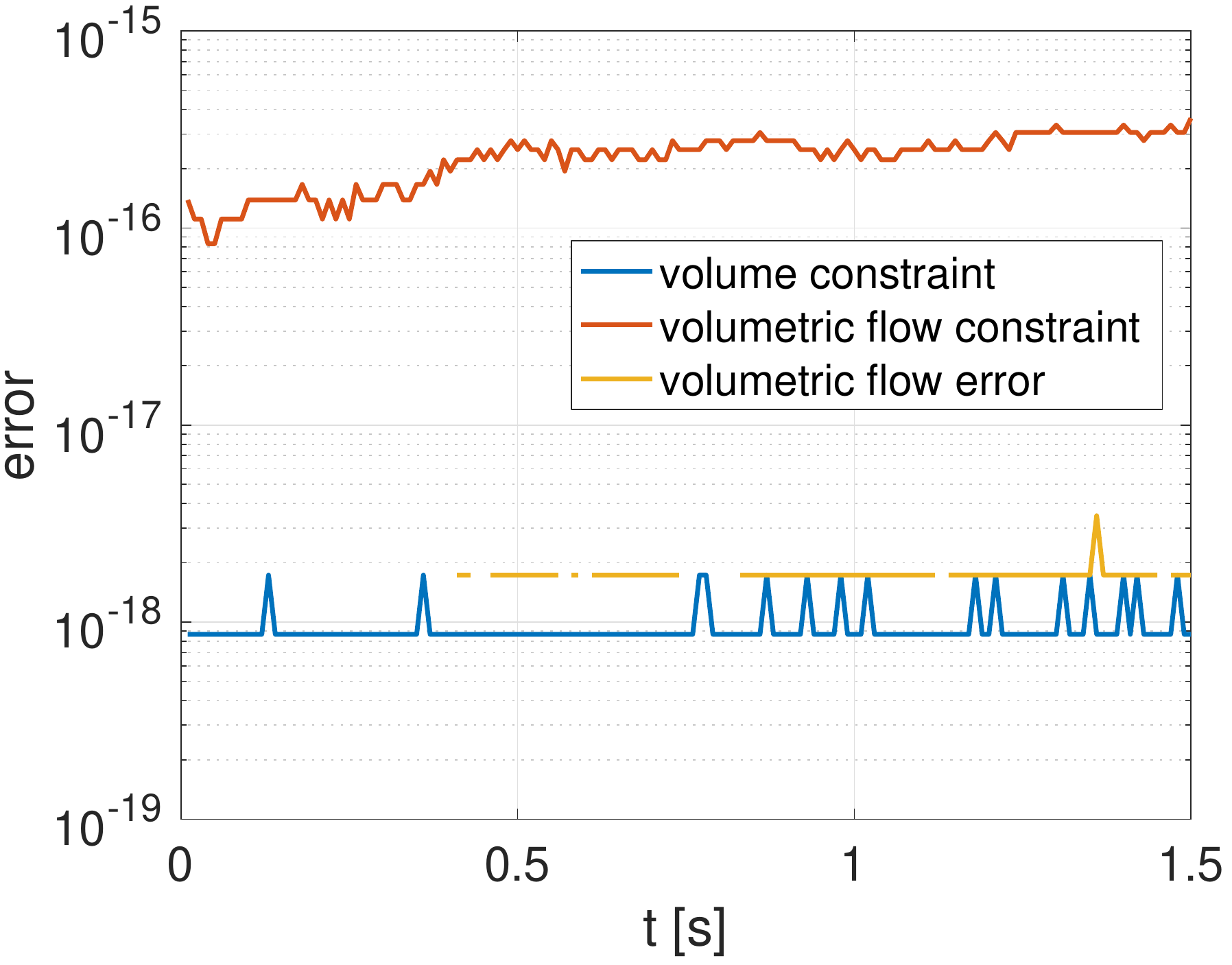}
\caption{Errors in the volume constraint, the volumetric flow constraint and the volumetric flow remain at machine precision ($\Delta t = \SI{1/100}{s}$).\label{fig:constraint_errors_KH_INC_PFTFM_N_40_dt001}}
\end{figure}

Figure \ref{fig:fullsolution_KH_INC_PFTFM_N_40_dt001} shows the solution at $t=\SI{1.5}{s}$ computed with $N=40$ finite volumes and $\Delta t=\SI{1/100}{s}$. The amplitude has grown with respect to the initial condition due to the negative imaginary part in $\omega_{2}$. At the same time, non-linear effects are causing wave steepening, which will lead to shock formation at later times. Figure \ref{fig:convergence_KH_INC_PFTFM_N_40} shows the error in the solution upon time step refinement, where the error is computed with respect to a reference solution computed with $\Delta t = \SI{1e-4}{s}$. Figure \ref{fig:constraint_errors_KH_INC_PFTFM_N_40_dt001} shows that the errors related to the volume constraint, the volumetric flow constraint, and the actual volumetric flow are satisfied until machine precision. Note that these errors are defined at each time instant as:
\begin{alignat}{2}
&\text{volume constraint error:}  \qquad && \max | \Omega_{p}^{-1}  \left( \frac{\ul{U}_{1}^{n}}{\rho_g} + \frac{\ul{U}_{2}^{n}}{\rho_l} \right) - A|, \label{eqn:error1}\\
&\text{volumetric flow constraint error:}  \qquad && \max | \Omega_{p}^{-1} D_{p} \Omega_{u}^{-1} \left( \frac{ \ul{U}_{3}^{n}}{\rho_g} + \frac{\ul{U}_{4}^{n}}{\rho_l} \right) |, \label{eqn:error2} \\
&\text{volumetric flow error:}  \qquad && \max | \Omega_{u}^{-1} \left( \frac{ \ul{U}_{3}^{n}}{\rho_g} + \frac{\ul{U}_{4}^{n}}{\rho_l} \right)  - \ul{Q}(t^{n}) |, \label{eqn:error3}
\end{alignat}
and the maximum is taken over all elements in the vector.

\FloatBarrier

\subsection{Discontinuous solutions: roll waves}
In this second test case we simulate roll waves on a periodic domain in order to (i) investigate the difference in $Q(t)$ between the proposed PF-TFM and the original TFM for the case of periodic boundary conditions, (ii) show that the PF-TFM captures the same discontinuous solution as the TFM, and (iii) give an indication of the computational speed-up that can be achieved with the PF-TFM. 

Roll waves can form when simulating the Kelvin-Helmholtz instability of section \ref{sec:KH} further into the nonlinear regime. The test case that we perform here is inspired by the experimental work of Johnson \cite{Johnson2005} and simulations of Akselsen \cite{Akselsen2018}, Holm\aa s \cite{Holmas2010}, and Sanderse \cite{Sanderse2018b}. The parameter settings are reported in table \ref{tab:Johnson}. We first compute a steady state solution: we prescribe the flow rates $Q_{g}/A = \SI{3.5}{m/s}$, $Q_{l}/A = \SI{0.35}{m/s}$, which leads to $\alpha_{l} = 0.190$ and $\frac{\rd p_{0}}{\rd s} = \SI{-155.919}{Pa/m}$. The initial condition follows by applying a sinusoidal perturbation of the form \eqref{eqn:analytical_linearized}, with a hold-up amplitude of 0.01, $k=2\pi / L$ and $\omega =4.597- 0.068 i$ (unit \si{1/s}) chosen such that a single, unstable, wave is triggered (similar to the Kelvin-Helmholtz test case in section \ref{sec:KH}). One important difference with respect to the previous test case is that the interfacial friction factor $f_{gl}$ in the expression for the interfacial stress ($\tau_{gl} = \frac{1}{2} f_{gl} \rho_{g} |u_{g}-u_{l}| (u_{g}-u_{l})$) is taken as $f_{gl} = m \cdot f_{g}$, with $m=12.5$. Another important difference is that the discretization of the convective terms in the momentum equation is performed with a first-order upwind scheme in order to prevent numerical oscillations at the discontinuity (shock wave), instead of the central scheme used in the first test case.

\begin{figure}[h!]
\centering
\begin{subfigure}[b]{.49\textwidth}
	\centering
\includegraphics[width=\textwidth]{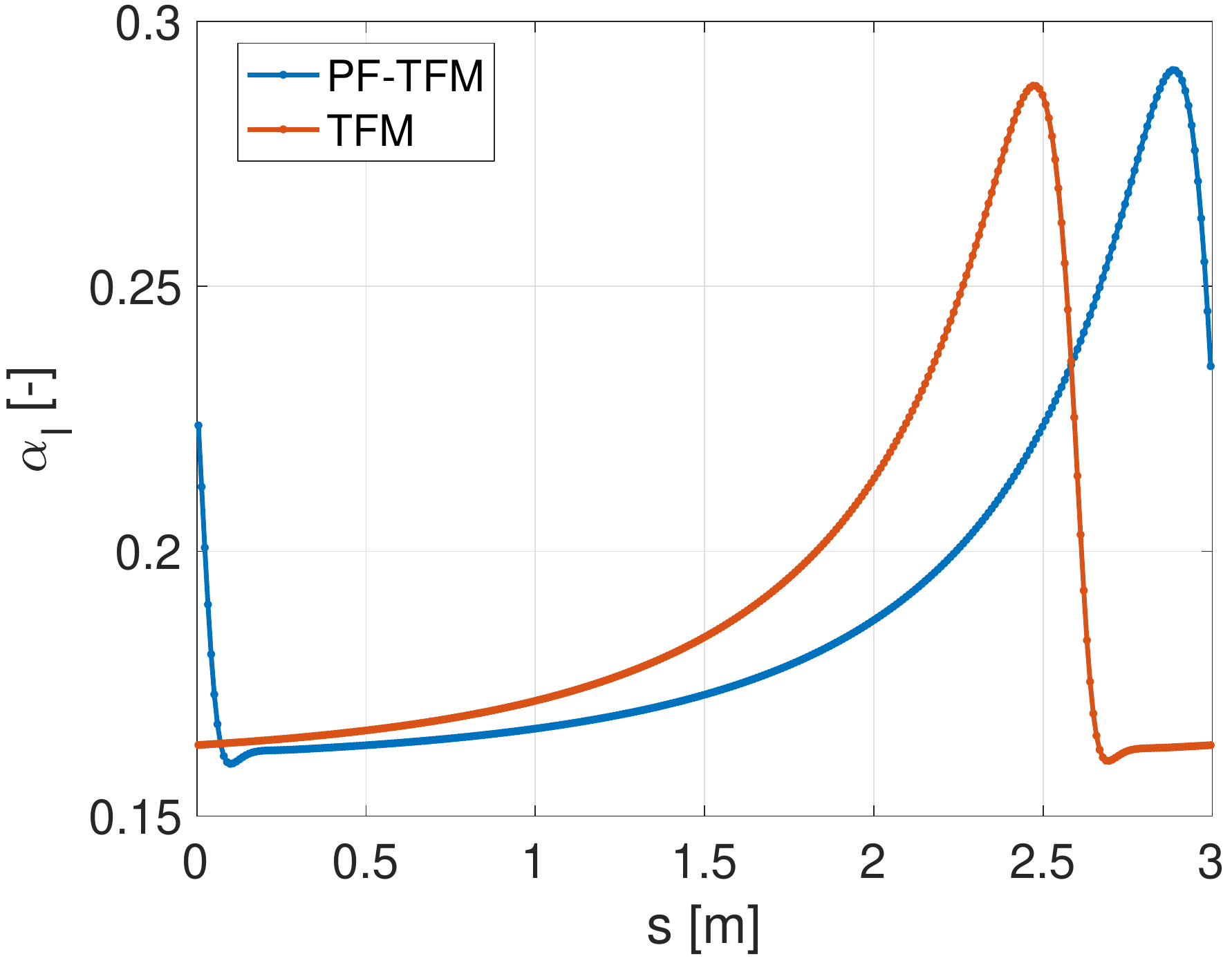}
\caption{Hold-up profiles at $t=\SI{100}{s}$.\label{fig:RW_holdup_IFE}}
	\end{subfigure}
	\hfill
\begin{subfigure}[b]{.49\textwidth}
	\centering
\includegraphics[width=\textwidth]{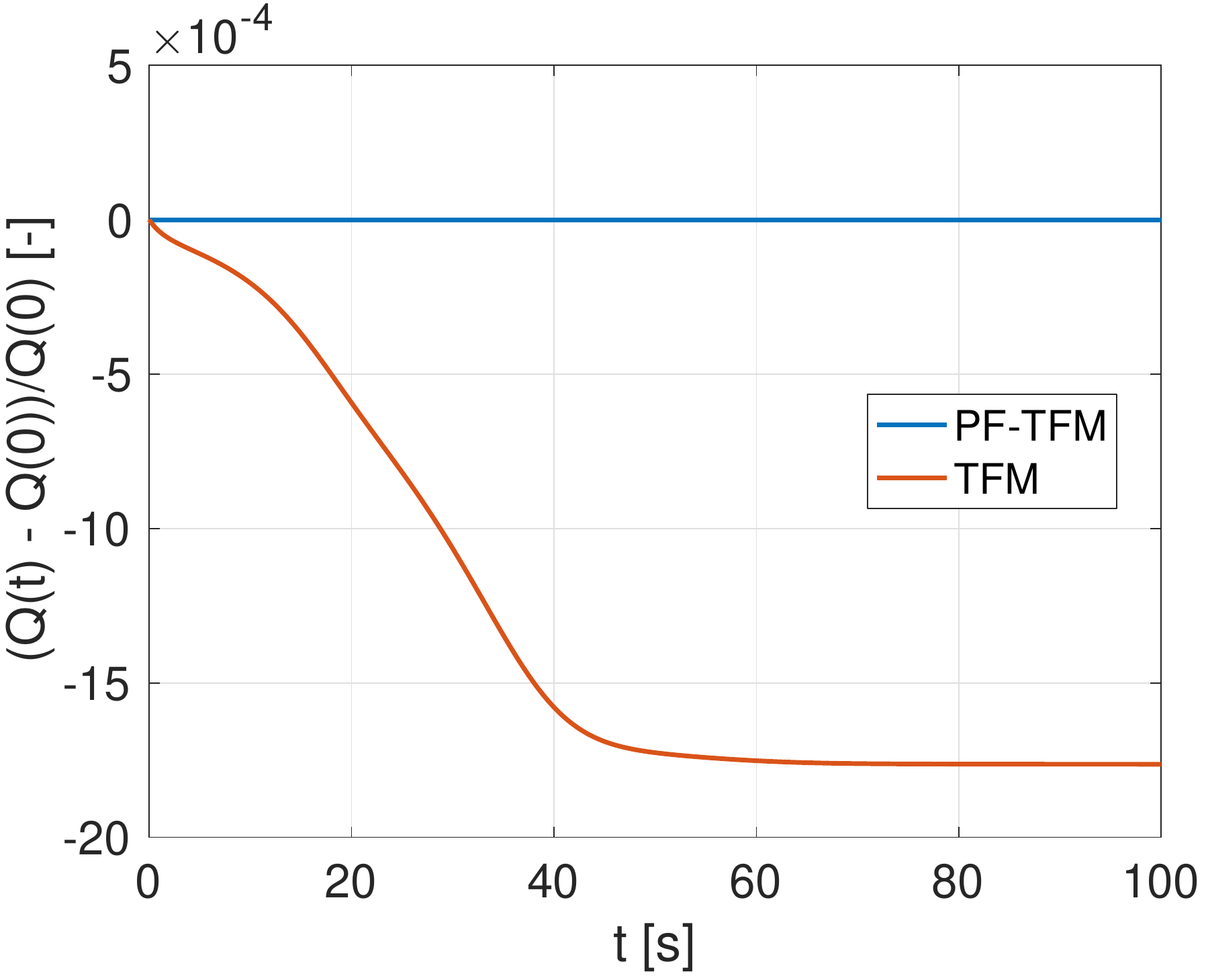}
\caption{Change in $Q(t)$ with respect to initial condition.\label{fig:RW_Qtot_IFE}}
	\end{subfigure}
\caption{Roll wave solutions computed with PF-TFM and TFM, with $N=320$ and $\Delta t=\SI{1/320}{s}$. \label{fig:RW}}
\end{figure}

Figure \ref{fig:RW_holdup_IFE} shows the hold-up profile at $t=\SI{100}{s}$ computed with both the PF-TFM and the TFM. At this time instant the wave has travelled approximately 73 times through the domain, covering a distance of about 219 m. The seemingly large difference in the position of the discontinuity (about 0.4 m) between the PF-TFM prediction and the TFM prediction is therefore in fact less than 0.2\% of the total travelled distance. This difference is consistent with the relative difference in the volumetric flow rate predicted by the PF-TFM and TFM, which is also approximately 0.2\% (see figure \ref{fig:RW_Qtot_IFE}). This difference reaches a constant value after approximately 50 seconds, as the roll wave then reaches a stationary solution. Note that the TFM prediction, which does not require a prescribed value of $Q(t)$, should be interpreted as being the most accurate solution. The PF-TFM prediction can be seen as a highly accurate approximation to this solution at a much reduced computational cost. \R{com5}\hl{The difference between the two solutions does not vanish upon grid or time-step refinement, since it is caused by a model assumption (namely constant volumetric flow) rather than a discretization effect. Fortunately, this assumption is only necessary for the case of periodic boundary conditions, as for other types of boundary conditions, $Q$ is known from the boundary condition values.}

\begin{figure}[h!]
	\centering
	\begin{subfigure}[b]{.49\textwidth}
\includegraphics[width=\textwidth]{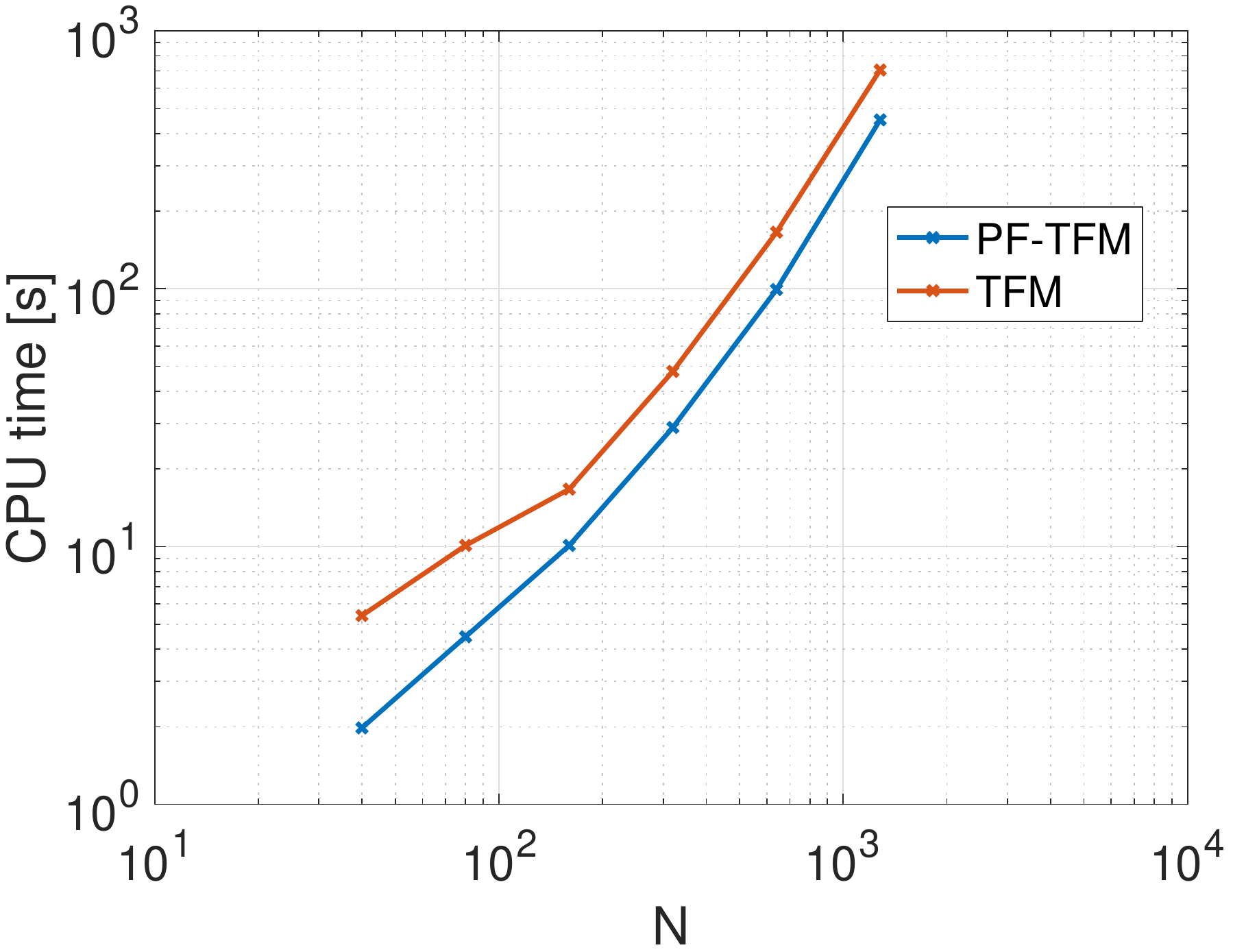}
\caption{Absolute CPU time.\label{fig:RW_CPU_IFE}}
\end{subfigure}
\hfill
\begin{subfigure}[b]{.49\textwidth}
\includegraphics[width=\textwidth]{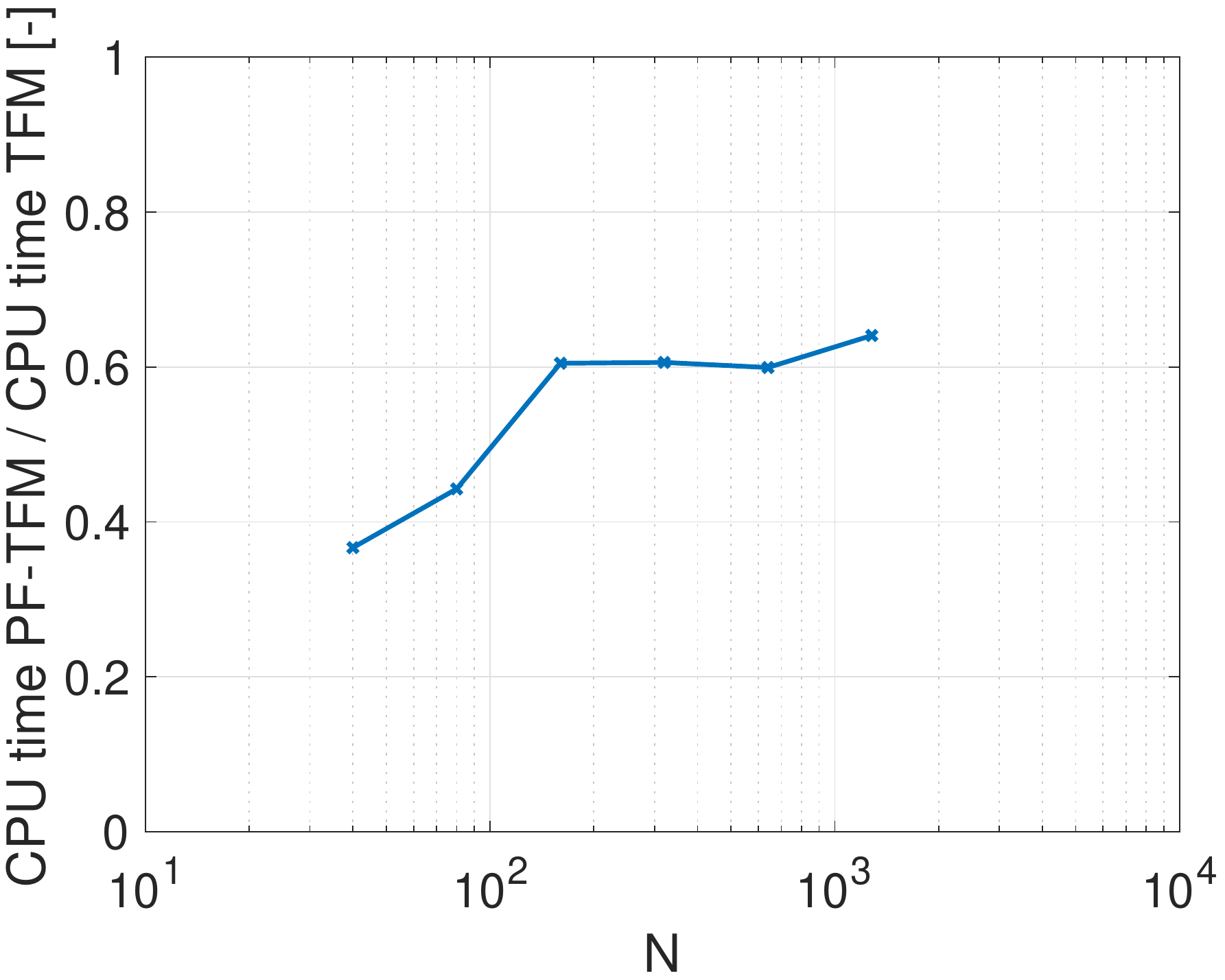}
\caption{Effective speed-up.\label{fig:RW_CPU_IFE_relative}}
\end{subfigure}
\caption{Comparison of computational time of PF-TFM and TFM.\label{fig:RW_CPU}}
\end{figure}

For this particular case ($N=320$), the computational cost of the PF-TFM prediction is about 40\% lower than of the TFM. For other grid sizes a similar cost reduction is observed, as is shown in figure \ref{fig:RW_CPU}\hl{, while the difference in accuracy remains negligible}. Figure \ref{fig:RW_CPU_IFE} shows a quadratic scaling of CPU time with $N$, which is caused by the fact that the simulations are run at a fixed CFL number ($\Delta t = 1/N$). Figure \ref{fig:RW_CPU_IFE_relative} shows that the obtained reduction in computational cost is relatively independent of the number of grid cells. However, we should note that other factors can have an important influence, such as the type of pressure solver used in the TFM (here a direct solver was used) or the type of friction model. For example, when a more computationally expensive friction model like the Biberg model \cite{Biberg2007} is used, the gain in computational cost obtained with the PF-TFM compared to the TFM is probably less pronounced, because the contribution of the pressure Poisson solve to the total computational cost is smaller. On the other hand, the computations are performed here on a single CPU. Much larger speed-ups with the PF-TFM can be expected in a parallel implementation: the PF-TFM framework is fully explicit and can be easily parallelized, whereas the original TFM has an implicit component (the solution of a Poisson equation) for which good parallel scaling will be more difficult to obtain.

\begin{table}[hbtp]
\centering
\caption{Parameter values for roll wave simulation. \label{tab:Johnson}}
\begin{tabular}{lrl}
\toprule
parameter & value & unit \\
\midrule
$\rho_{l}$  & $998$ & \si{kg/m^{3}} \\ 
$\rho_{g}$  & $50$ & \si{kg/m^{3}} \\ 
$R$ & $0.05$ & \si{m} \\
$g$ & $9.8$ & \si{m/s^{2}} \\
$\mu_{g}$ & $1.61 \cdot 10^{-5}$ & \si{Pa.s} \\
$\mu_{l}$ & $1 \cdot 10^{-3}$ & \si{Pa.s} \\
$\varphi$ & $0$ & \si{deg} \\
$\epsilon$ & $2\cdot 10^{-5}$ & \si{m} \\
$L$ & $3$ & \si{m} \\
\bottomrule
\end{tabular}
\end{table}

\FloatBarrier

\subsection{Hold-up wave propagation}
The final test case considers the propagation of a hold-up wave through a horizontal pipeline caused by an increasing inlet gas flow rate, inspired by the test case in \cite{Omgba-Essama2004}. The parameters of the problem are described in table \ref{tab:IFP_parameters}. The initial conditions are steady state production with inlet mass flows of liquid and gas of $I_{l}=\SI{1}{kg/s}$ and $I_{g,\text{start}}=\SI{0.02}{kg/s}$. Furthermore, we specify the following time-dependent gas flow rate at the inlet,
\begin{equation}
%I_{g} (t) = I_{g,\text{start}}  + \frac{1}{2}( I_{g,\text{end}}- I_{g,\text{start}} ) (1 + \tanh( (t-t_{\text{jump}})/ \Delta_{\text{jump}})
I_{g}(t) = I_{g,\text{start}}  + ( I_{g,\text{end}}- I_{g,\text{start}} ) e^{1/2-200/t} /e^{1/2},
\end{equation}
which is such that $\dot{Q}(0)=0$ and for $t>0$ $\dot{Q}(t)$ (and higher order derivatives) are nonzero.
The volumetric flows $Q_{g}(t)$ and $Q_{l}$ are given by $I_{g}(t)/\rho_{g}$ and $I_{l}/\rho_{l}$, respectively.

\begin{table}[hbtp]
\centering
\caption{Parameter values for the IFP problem. \label{tab:IFP_parameters}}
\begin{tabular}{lrl}
\toprule
parameter & value & unit \\
\midrule
$\rho_{l}$  & $1003$ & \si{kg/m^{3}} \\ 
$\rho_{g}$  & $1.26$ & \si{kg/m^{3}} \\ 
$R$ & $0.073$ & \si{m} \\
$g$ & $9.8$ & \si{m/s^{2}} \\
$\mu_{g}$ & $1.8 \cdot 10^{-5}$ & \si{Pa.s} \\
$\mu_{l}$ & $1.516 \cdot 10^{-3}$ & \si{Pa.s} \\ 
$\epsilon$ & $10^{-8}$ & \si{m} \\
$L$ & $1000$ & \si{m} \\
\bottomrule
\end{tabular}
\end{table}

\begin{figure}[h!]
\centering
\begin{subfigure}[b]{.49\textwidth}
	\centering
\includegraphics[width=\textwidth]{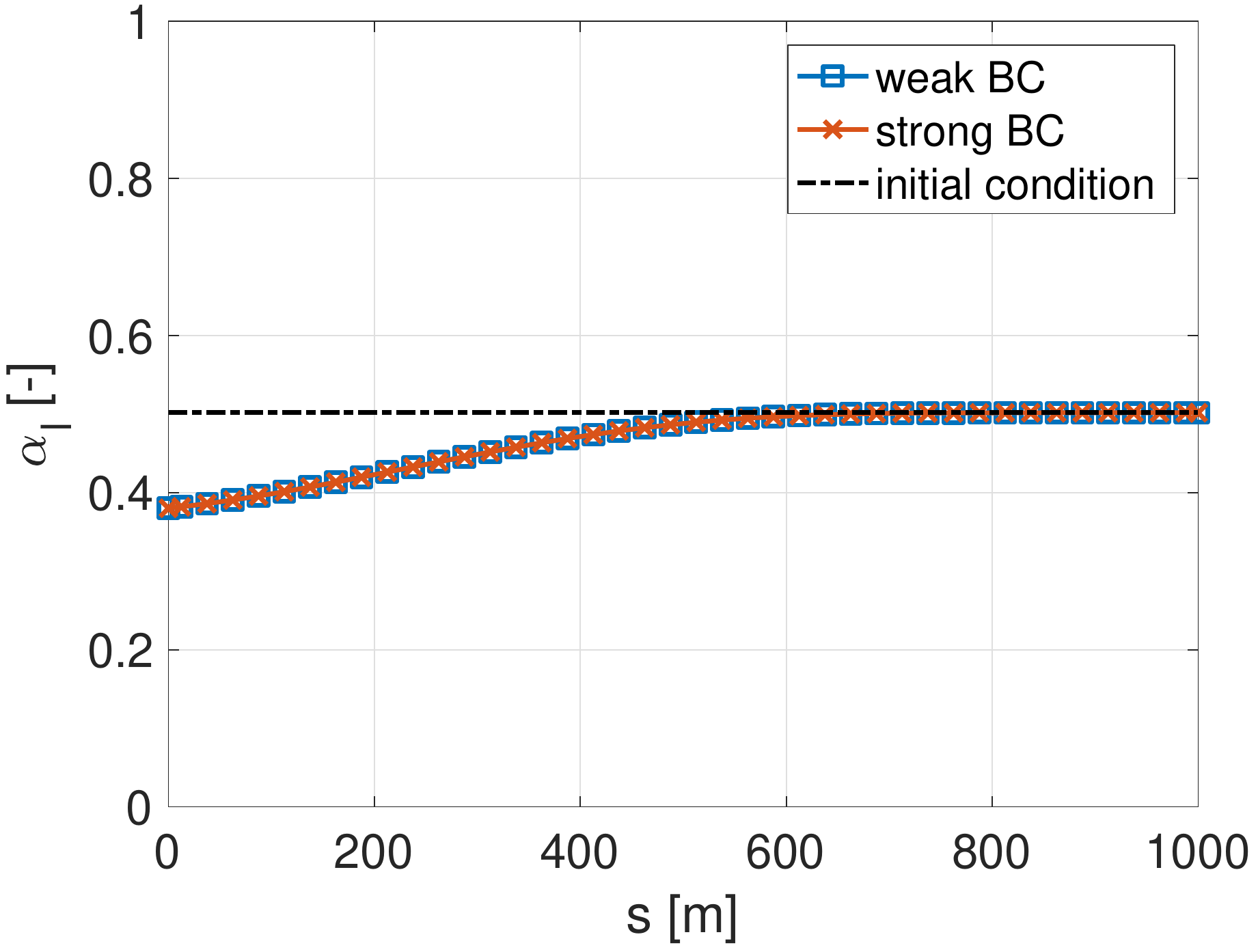}
\caption{Solution profile at $t=\SI{1000}{s}$.\label{fig:holdup_IFP_t1000}}
	\end{subfigure}
	\hfill
\begin{subfigure}[b]{.49\textwidth}
	\centering
\includegraphics[width=\textwidth]{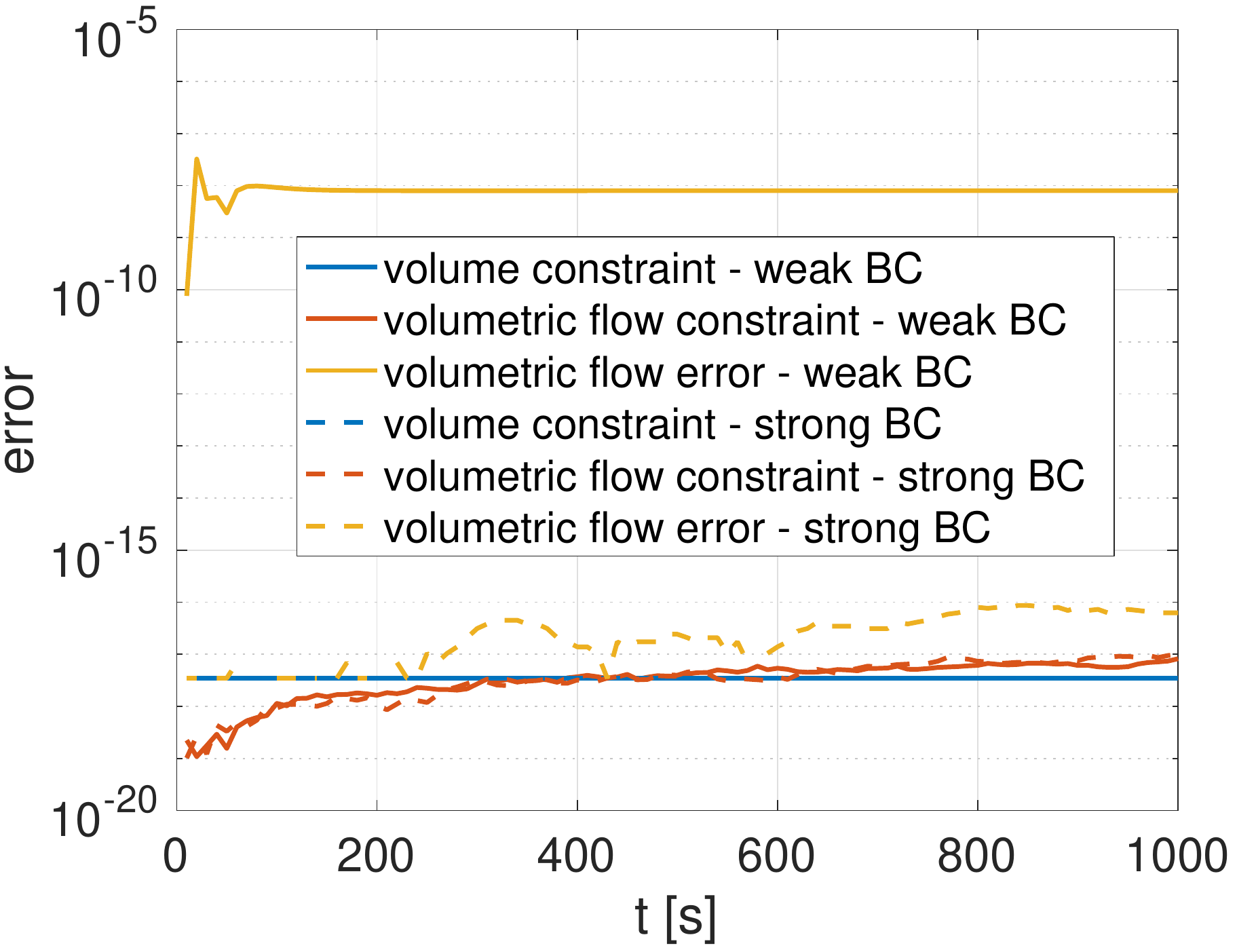}
\caption{Constraint errors as a function of time.\label{fig:constraints_weak_strong_dt10}}
	\end{subfigure}
\caption{Hold-up wave propagation with $N=40$ and $\Delta t=\SI{10}{s}$. The weak BC corresponds to option A, the strong BC to option B.\label{fig:hold-upwave}}
\end{figure}

The solutions with the weak BC prescription (option A) and with the strong BC prescription (option B) at $t=\SI{1000}{s}$ with $N=40$ and $\Delta t = \SI{10}{s}$ are indistinguishable, see figure \ref{fig:holdup_IFP_t1000}. Figure \ref{fig:constraints_weak_strong_dt10} shows that the errors in volume constraint and volumetric flow constraint (equations \eqref{eqn:error1}-\eqref{eqn:error2}) remain at machine precision for both methods. In addition, option B has the property that the numerically computed volumetric flow rate remains exactly equal to the specified value. In option A, a small error is present, which decreases with third order upon time step refinement.

\begin{figure}[h!]
\centering
\begin{subfigure}[b]{.49\textwidth}
	\centering
\includegraphics[width=\textwidth]{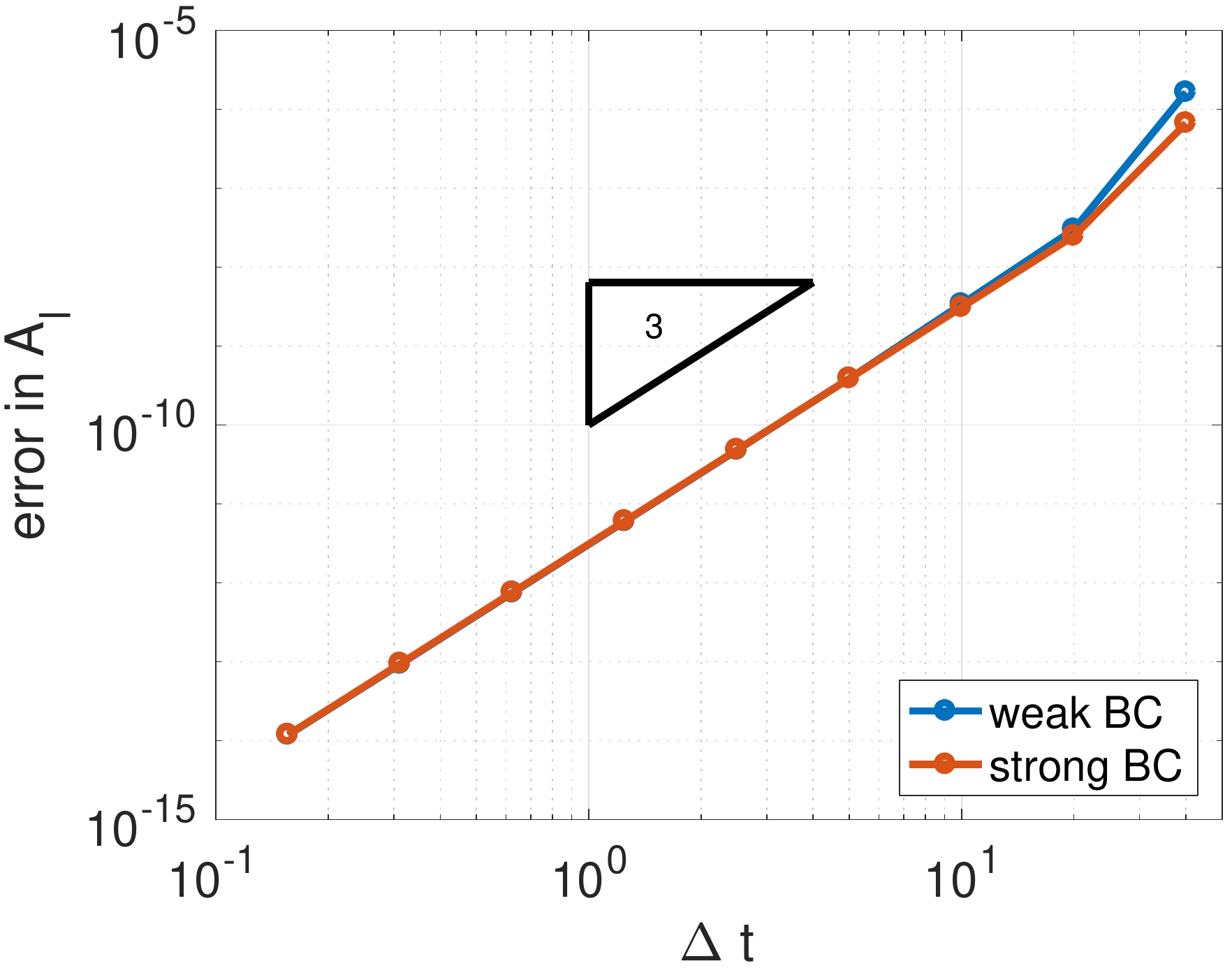}
\caption{Error in hold-up.\label{fig:convergence_IFP_PFTFM_INC_N_40_A_l}}
	\end{subfigure}
	\hfill
\begin{subfigure}[b]{.49\textwidth}
	\centering
\includegraphics[width=\textwidth]{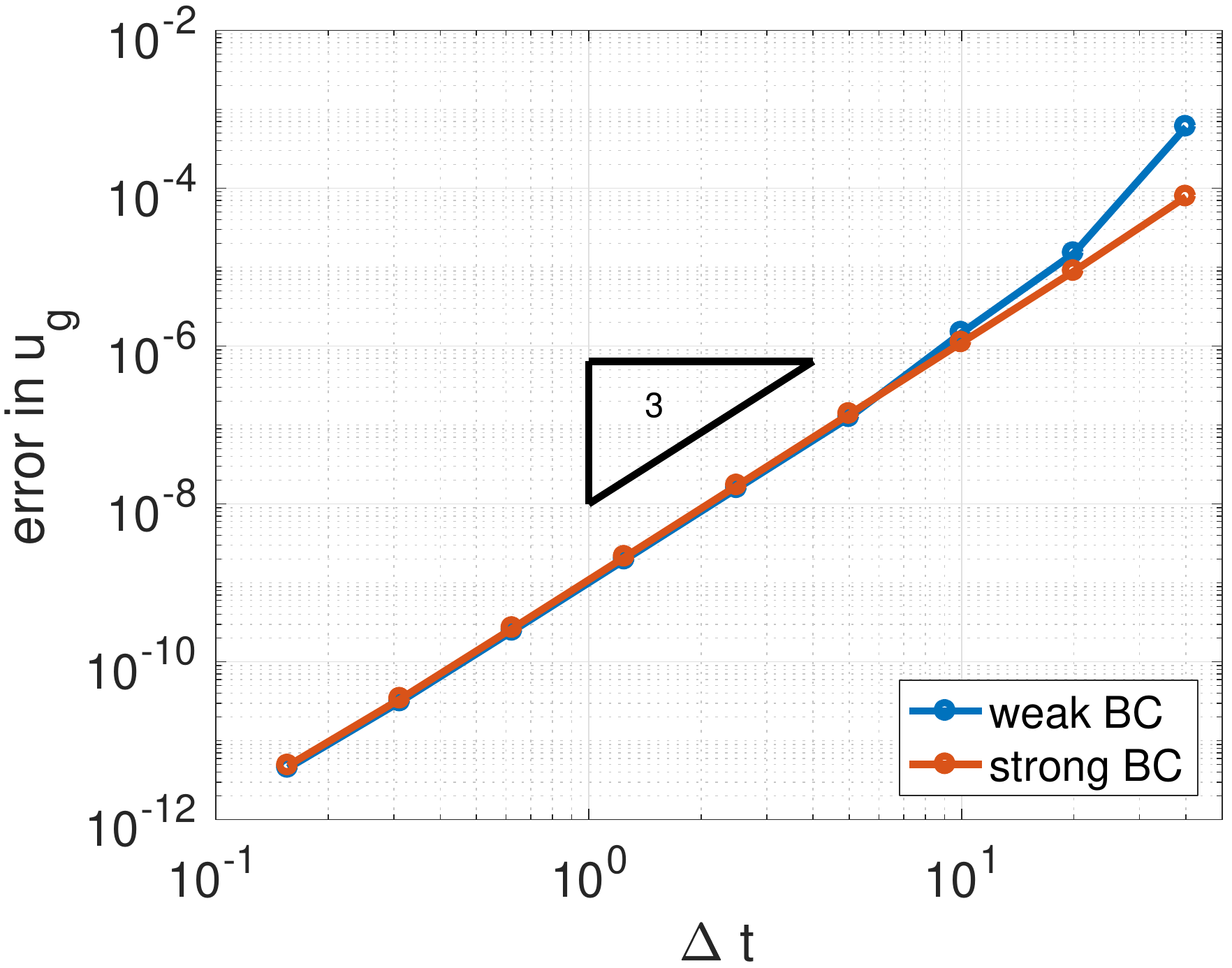}
\caption{Error in gas velocity. \label{fig:convergence_IFP_PFTFM_INC_N_40_u_g}}
	\end{subfigure}
\caption{Third-order convergence of the error at $t=\SI{1000}{s}$ for hold-up wave propagation. The weak BC corresponds to option A, the strong BC to option B.\label{fig:convergence_IFP_PFTFM_INC_N_40}}
\end{figure}

A more quantitative assessment is obtained by computing a reference solution at $t=\SI{1000}{s}$ with $N=40$, $\Delta t= \SI{1e-2}{s}$, and weak boundary conditions (option A). At this small time step, the volumetric flow error stays at machine precision also with weak imposition of $Q$ and boundary conditions. Next we compute solutions at a sequence of much larger time steps, $\Delta t=40, 20, 10, \ldots$, both with weak BC (option A) and strong BC (option B). Figure \ref{fig:convergence_IFP_PFTFM_INC_N_40} shows that both options converge according to the design order of accuracy (third order).

\FloatBarrier

%Option A vs. option B. We compute a reference solution with option A with a small time step, such that the volumetric flow error is at machine precision. We compute errors with respect to this reference solution. We find correct convergence when comparing with option A results, but not when comparing with option B for fine time steps. Possible causes:
%\begin{itemize}
%\item The derivation of the RK scheme for option B is wrong. We checked a 2nd order scheme, which converges fine, but does not get to the accuracy where the third order scheme stops converging.
%\item The option A solution is not accurate enough. \textit{Probably not, the convergence of option A looks fine}
%\item There is a bug, e.g.\ a timelevel misspecification when calling $I$ or $dI/dt$.
%\item There is a postprocessing step that changes the solution. \textit{Unlikely, this is only done if the volumetric flow is not correct.}
%\item The adaptation of the initial solution to make it divergence free and equal to specified $Q$ spoils the accuracy. \textit{Probably not, the adaptation of the initial solution is not time-step dependent.}
%\end{itemize}

\section{Conclusions}
In this article we have proposed a novel pressure-free incompressible two-fluid model for multiphase pipeline transport, in which both the pressure and the volume constraint have been eliminated. Together with a staggered-grid spatial discretization and an explicit Runge-Kutta time integration method, this yields a new highly efficient method for solving the incompressible two-fluid model equations. The explicit nature also allows for a straightforward parallelization. Furthermore, the absence of the pressure make\R{com7}\hl{s} the model more amenable to analysis. For example, the PF-TFM  offers a new viewpoint in studying the issue of the conditional well-posedness of the TFM. 
%This conditional well-posedness of the TFM is not the subject of this paper, but the proposed alternative formulation of the TFM developed in this paper is expected to aid the study of the well-posedness issue from a new angle.
The `price' to pay compared to the conventional form of the incompressible two-fluid model, is that the time-derivative of the volumetric flow rate needs to be specified as an additional input to the equations. Fortunately, in most cases for pipeline transport design (e.g.\ steady production, production ramp-up) this value is known from the boundary conditions.

A natural question is whether the original TFM `pressure Poisson' formulation \cite{Sanderse2018} is still useful, given that a more efficient pressure-free alternative has been proposed in this work. The answer is yes, for two reasons. First, the original formulation also works when $\dot{Q}$ is unknown (e.g.\ in the case of periodic boundary conditions). Second, as indicated in \cite{Sanderse2018}, the original formulation has the potential to be extended to multi-dimensional problems, with application to for example the volume-of-fluid approach. This is probably not possible with the proposed pressure-free approach: the fact that the pressure gradient can be expressed directly in terms of the conservative variables by using the constraint is presumably only possible when dealing with one-dimensional problems. 

Extension of the pressure-free model to one-dimensional incompressible flow with three instead of two fluid phases 	is straightforward. \R{com3_2}\hl{Extension with an energy equation is also likely to be possible, since in incompressible flow the energy equation decouples from the mass and momentum equations (it can be solved once the mass and momentum equations have been solved).} On the other hand, extension to compressible flow is not straightforward: in compressible flow, differentiation of the volume constraint leads to a pressure equation that involves the time derivative of the pressure. Consequently, the pressure gradient term in the momentum equations cannot be eliminated in the same way as in the incompressible model.

\section*{CRediT}
\textbf{Benjamin Sanderse}: Conceptualization, Methodology, Software, Writing - Original Draft; \textbf{Jurriaan Buist}: Writing - Review \& Editing; \textbf{Ruud Henkes}: Resources,  Writing - Review \& Editing.

\FloatBarrier

\appendix % Reset the chapter name and the chapter counting.

\section{Eigenvalues of the PF-TFM}\label{sec:eigenstructure}
To investigate the eigenvalues of the PF-TFM, one studies the terms in the model that involve partial derivatives, i.e.\
\begin{equation}
\frac{\partial \vt{U}}{\partial t} + \vt{A} (\vt{U}) \frac{\partial \vt{f}(\vt{U})}{\partial s}. %= \frac{\partial \vt{U}}{\partial t} + \vt{A} (\vt{U}) \dd{\vt{f}}{\vt{U}}  \dd{\vt{U}}{s}.
\end{equation}
We are thus interested in the eigenvalues of the matrix
\begin{equation}
\vt{A} (\vt{U}) \dd{\vt{f}(\vt{U})}{\vt{U}} =
\begin{pmatrix}
0 & 0& 1& 0 \\
0 & 0& 0& 1 \\
- A_{l}  \rho_{g} (A_{g} g + u_{g}^2)/ \hat{\rho}& - A_{g} \rho_{g}  (A_l g - u_{l}^2) / \hat{\rho} &  2 A_{l}  \rho_{g} u_{g} / \hat{\rho} & - 2 A_{g} \rho_{g} u_{l} / \hat{\rho} \\
 A_{l}  \rho_{l} (A_{g} g + u_{g}^2)/ \hat{\rho} &  A_{g} \rho_{l}  (A_l g - u_{l}^2) /\hat{\rho}&  -2 A_{l}  \rho_{l} u_{g} / \hat{\rho} &  2 A_{g} \rho_{l} u_{l} / \hat{\rho}
\end{pmatrix}.
\end{equation}
These eigenvalues read
\begin{equation}
\lambda_{1,2} = \frac{A_{l} \rho_{g} u_{g} + A_{g} \rho_{l} u_{l} \pm \sqrt{A_{g} A_{l} \left( g \hat{\rho} \Delta \rho - \rho_{g} \rho_{l} (\Delta u)^2\right)}}{\hat{\rho}},
\end{equation}
where $\Delta \rho = \rho_{l} - \rho_{g}$ and $\Delta u = u_{l} - u_{g}$, and
\begin{equation}
\lambda_{3,4} = 0.
\end{equation}
These eigenvalues are the same as those of the TFM, see e.g.\ \cite{Sanderse2018}.
%
%The corresponding eigenvectors are...
%\todo{add eigenvectors}

\section{Pressure-free model from fully discrete two-fluid model equations}\label{sec:PF_fullydiscrete}
In this appendix we derive a discrete PF-TFM based on a fully discrete approximation of the TFM. This corresponds to `option B'. The forward Euler discretization of the TFM reads \cite{Sanderse2018}:
\begin{align}\label{eqn:FE_TFM}
\ul{U}_{12}^{n+1} &= \ul{U}_{12}^{n}  +  \Delta t \ul{F}_{\text{TFM},12} (\ul{U}^{n}), \\
\ul{U}_{34}^{n+1} &= \ul{U}_{34}^{n}  +  \Delta t \ul{F}_{\text{TFM},34} (\ul{U}^{n}) - \Delta t H(\ul{U}^{n}) p^{n}, \\ 
\frac{\Omega_{p}^{-1} \ul{U}_{1}^{n+1}}{\rho_{g}} + \frac{\Omega_{p}^{-1}  \ul{U}_{2}^{n+1}}{\rho_{l}} &= A,
\end{align}
where $H$ denotes the discretized pressure gradient operator, with the two components
\begin{equation}
H_{3} (\ul{U}) p= ( I_{p} \ul{U}_{1}/\rho_{g}) \odot (G p), \qquad H_{4} (\ul{U}) p= (I_{p} \ul{U}_{2}/\rho_{l}) \odot (G p).
\end{equation}
 $I_{p} \in \mathbb{R}^{N_u \times N_p}$ is an interpolation matrix, $G \in \mathbb{R}^{N_u \times N_p}$ a differencing matrix (similar to $D_p$), and $\odot$ denotes elementwise multiplication.
To arrive at the pressure-free model based on the discrete two-fluid model, we follow exactly the same steps as in the continuous case, outlined in section \ref{sec:PFTFM}. We substitute the mass equations in the constraint and apply the expression for $\ul{F}_{\text{TFM},12}$:
\begin{equation}
\Omega_{p}^{-1}  \left( \frac{ \ul{U}_{1}^{n}}{\rho_{g}} + \frac{ \ul{U}_{2}^{n}}{\rho_{l}}\right)  +  \Delta t \Omega_{p}^{-1} D_{p} \Omega_{u}^{-1} \left( \frac{\ul{U}_{3}^{n}}{\rho_{g}} + \frac{ \ul{U}_{4}^{n} }{\rho_{l}} \right) = A,
%M \ul{U}_{12}^{n}  +  \Delta t M D_p \Omega_{u}^{-1} \ul{U}_{34}^{n} = 0.
\end{equation}
which reduces to
\begin{equation}
\Omega_{p}^{-1} D_{p} \Omega_{u}^{-1} \left( \frac{\ul{U}_{3}^{n}}{\rho_{g}} + \frac{ \ul{U}_{4}^{n} }{\rho_{l}} \right) = 0.
\end{equation}
$D_p$ is a differencing matrix with two diagonals ($-1$ and $1$), whose nullspace consists of a constant vector. Consequently, we have 
\begin{equation}
\Omega_{u}^{-1} \left( \frac{\ul{U}_{3}^{n}}{\rho_{g}}  + \frac{\ul{U}_{4}^{n}}{\rho_{l}} \right) = k \ul{1},
\end{equation}
where $\ul{1} \in \mathbb{R}^{N_{u}}$ is a vector with ones as entries and the constant $k$ will be \textit{chosen} to equal $Q(t^n)$. Inserting the equation for the momenta from the previous time step and rewriting as
\begin{equation}
\frac{\ul{F}_{\text{TFM},3} (\ul{U}^{n-1})}{\rho_{g}} + \frac{\ul{F}_{\text{TFM},4} (\ul{U}^{n-1})}{\rho_{l}} -  \left( \frac{H_{3}(\ul{U}^{n-1})}{\rho_{g}} +  \frac{H_{4}(\ul{U}^{n-1})}{\rho_{l}}  \right) p^{n-1}  = \Omega_{u} \frac{\ul{Q}^{n}  - \ul{Q}^{n-1}}{\Delta t},
\end{equation}
leads to
\begin{equation}
G p^{n-1}  = \frac{\rho_{l} \rho_{g}}{\ul{\hat{\rho}}^{n-1} } \odot \left( \frac{\ul{F}_{\text{TFM},3} (\ul{U}^{n-1})}{\rho_{g}} + \frac{\ul{F}_{\text{TFM},4} (\ul{U}^{n-1})}{\rho_{l}} - \Omega_{u} \frac{\ul{Q}^{n}  - \ul{Q}^{n-1}}{\Delta t} \right).
\end{equation}
We have thus obtained the pressure equation by rewriting the fully discretized TFM and additionally supplying knowledge of the volumetric flow $Q^{n}$. Note that the division by $\ul{\hat{\rho}}$ is elementwise, and 
\begin{equation}
\ul{\hat{\rho}} = I_{p} \left( \frac{\ul{U}_{1}}{\rho_{g}} \rho_{l} + \frac{\ul{U}_{2}}{\rho_{l}} \rho_{g}  \right).
\end{equation}
%\begin{equation}
%(I_{p} ( U_{1}/\rho_{g}^{2} + U_{2}/\rho_{l}^{2} )) \odot G p^{n-2}  = \ul{F}_{\text{TFM},34} (\ul{U}^{n-1}) + \frac{Q^{n}  -Q^{n-1}}{\Delta t} \ul{1}   
%\end{equation}
%This equation can be solved for the pressure:
%\begin{equation}
%p^{n-2}  = \ul{U}_{34}^{n-1}  +  \Delta t \ul{F}_{\text{TFM},34} (\ul{U}^{n-1}) - Q^{n} \ul{1}.
%\end{equation}
Substituting the expression for the pressure gradient back into the momentum equation gives
\begin{equation}
\begin{split}
\ul{U}_{3}^{n+1} &= \ul{U}_{3}^{n}  + \Delta t \ul{F}_{3} (\ul{U}^{n}), 
\end{split}
\end{equation}
%\begin{equation}
%\begin{split}
%\ul{U}_{4}^{n+1} &= \ul{U}_{34}^{n}  +  \Delta t \ul{F}_{\text{TFM},34} (\ul{U}^{n}) - \Delta t \left( \ul{U}_{34}^{n-1}  +  \Delta t \ul{F}_{\text{TFM},34} (\ul{U}^{n-1}) - Q^{n} \ul{1} \right) 
%\end{split}
%\end{equation}
where
\begin{equation}\label{eqn:F3}
\begin{split}
\ul{F}_{3} (\ul{U}^{n}) =& \underbrace{\left( \ul{1} -  \frac{I_{p} \ul{U}_{1}^{n}}{\rho_{g}} \odot \frac{\rho_{l}}{\ul{\hat{\rho}}^{n}} \right) \odot \ul{F}_{\text{TFM},3} (\ul{U}^{n}) - \left( \frac{I_{p} \ul{U}_{1}^{n}}{\rho_{g}} \odot \frac{ \rho_{g}}{\ul{\hat{\rho}}^{n}} \right) \odot \ul{F}_{\text{TFM},4} (\ul{U}^{n})}_{\ul{\hat{F}_{3}} (\ul{U}^{n}) } \\ 
& + \underbrace{\left( \frac{I_{p} \ul{U}_{1}^{n}}{\rho_{g}} \odot \frac{\rho_{l} \rho_{g}}{\ul{\hat{\rho}}^{n} } \right) \odot \left( \Omega_{u} \frac{\ul{Q}^{n}  - \ul{Q}^{n-1}}{\Delta t} \right)}_{\ul{c}_{3}(\ul{U}^{n}) (Q^{n} - Q^{n-1})/\Delta t}.
\end{split}
\end{equation}
Note that the first term in brackets $(.)$ corresponds to the matrix element $A_{33}$ and the second term in brackets to $A_{34}$. 

In summary, the PF-TFM forward Euler discretization derived from the fully discrete forward Euler discretization of the TFM reads
\begin{equation}
\ul{U}^{n+1} = \ul{U}^{n}  + \Delta t \ul{\hat{F}} (\ul{U}^{n})  + \ul{c}(\ul{U}^{n}) (Q^{n} - Q^{n-1}),
\end{equation}
and is such that the volumetric flow stays exact when integrating in time (option B). The extension to a generic explicit Runge-Kutta method is a straightforward substitution exercise.

%\input{appendix_testcases}

%%%%%%%%%%%%%%%%%%      References     %%%%%%%%%%%%%%%%%%
\section*{\refname}
%\section{References}
\bibliography{extracted_literature}      % for an external bibliography file
\bibliographystyle{abbrv}

\end{document}